\documentclass[12pt]{article}
\usepackage{scicite}
\usepackage{graphicx}% Include figure files
\usepackage{amsmath}
\usepackage{times}
\usepackage{changes}
\usepackage{xcolor}
\usepackage{amssymb}
\usepackage{hyperref}

\title{Transition from Light Diffusion to Localization in Three-Dimensional Amorphous Dielectric Networks near the Band Edge}

\author
{Jakub Haberko,$^{1\ast}$ Luis S.\ Froufe-P\'erez,$^{2\ast}$ Frank Scheffold,$^{2\dagger}$\\
\\
\normalsize{$^{1}$Faculty of Physics and Applied Computer Science, AGH University}\\
\normalsize{of Science and Technology, al. Mickiewicza 30, 30-059 Krakow, Poland}\\
\normalsize{$^{2}$Department of Physics, University of Fribourg, 1700 Fribourg, Switzerland}\\
\\
\normalsize{$^\ast$ These authors contributed equally to this work.}
\\
\normalsize{$^\dagger$To whom correspondence should be addressed; E-mail:  frank.scheffold@unifr.ch.}
}

\date{}

\begin{document} 

% Double-space the manuscript.

\baselineskip24pt

% Make the title.

\maketitle

% Place your abstract within the special {sciabstract} environment.

%%%%%%%%%%%%%%%%%%%%%%%%%%%%%%%%%%%%%%%%%%%%%%%%%%%%%%%%%%%%%%%%%%%%%%%%%%%%%%%%%%%%%%%%%%
% INTRODUCTION
%%%%%%%%%%%%%%%%%%%%%%%%%%%%%%%%%%%%%%%%%%%%%%%%%%%%%%%%%%%%%%%%%%%%%%%%%%%%%%%%%%%%%%%%%%
\begin{abstract}
Localization of light is the photon analog of electron localization in disordered lattices for whose discovery Anderson received the Nobel prize in 1977. The question about its existence in open three-dimensional materials has eluded an experimental and full theoretical verification for decades. Here we study numerically electromagnetic vector wave transmittance through realistic digital representations of hyperuniform dielectric networks, a new class of highly correlated but disordered photonic band gap materials. We identify the evanescent decay of the transmitted power in the gap and diffusive transport far from the gap. Near the gap, we find that transport sets off diffusive but, with increasing slab thickness, crosses over gradually to a faster decay, signaling localization. We show that we can describe the transition to localization at the mobility edge using the self-consistent theory of localization based on the concept of a position-dependent diffusion coefficient. 
\end{abstract}
Band gap formation and strong Anderson localization (SAL) of classical waves are both considered general wave phenomena where the mechanism leading to the exponential attenuation of wave transport can be understood in terms of interference of scattered waves. In a periodically repeating environment, scattering of waves from Bragg planes can be associated with the opening of a photonic band gap (PBG) \cite{yablonovitch1987inhibited,john1987strong,Joannopoulos_book}.
\newline \indent The SAL mechanism is usually explained by the constructive interference of multiply scattered waves propagating along time-reversed loops, which increases the return probability and eventually leads to a breakdown of wave diffusion \cite{anderson1985question,van2000reflection,rotter2017light}. In contrast to PBG-formation the transition to SAL in disordered media strongly depends on dimensionality. In one and two dimensions, there are no truly extended states, and waves in disordered media are always localized for sufficiently large systems sizes \cite{anderson1958absence, chabanov2000statistical,froufe2007statistical,schwartz2007transport,conley2014light}. Only in three dimensions, wave localization shows a phase transition, and localized states appear below the 'mobility edge'. The threshold for localization can be estimated by the Ioffe-Regel criterion $k \cdot \ell \sim 1$, where $\ell$ denotes the transport mean free path and $k$ the wavenumber \cite{ioffe1960non,abrahams1979scaling,skipetrov2018ioffe}.  
The Anderson transition in three dimensions is a fascinating phenomenon that until now has eluded a full theoretical and experimental understanding. It's relevance is not restricted to transport of photons or electrons, but it can also be applied in acoustics, or any kind of coherent wave propagation \cite{hu2008localization}. Moreover, a full understanding and control over the wave fields in complex media offer a plethora of opportunities for applications in imaging, sensing, and photonics \cite{rotter2017light}.  However, an experimental observation of SAL for electromagnetic vector waves in three-dimensional systems has not yet been achieved, even though several claims to its existence were made \cite{wiersma1997localization,storzer2006observation,sperling2013direct} but soon after were put in question \cite{scheffold1999localization,scheffold2013inelastic} and later refuted \cite{skipetrov2016red}.  At about the same time it was found theoretically that SAL of electromagnetic vector waves is absent in random ensembles of point scatterers, irrespective of their scattering strength \cite{skipetrov2014absence}. Thus, in a 2016 perspective article, Skipetrov and Page declare a \emph{Red Light for Anderson localization} of light \cite{skipetrov2016red} and Maret et al. ask \cite{Maret_NJP_2016} \emph{whether 3D light localization can be reached in 'white paint'} at all?
\newline  \indent The advent of amorphous PBG materials over the last decade \cite{Edagawa2008,Florescu_PNAS_2009,Cao_PRA_2011,Froufe_PRL_2016,ricouvier2019foam,Klatt23480}, has opened a new pathway towards the design of strongly photonic dielectric materials. 
It has also prompted fundamental questions concerning the relationship between the Anderson localization transition and the transition to a full bandgap. Recently, we proposed a transport phase diagram for two-dimensional hyperuniform disordered PBG-materials with a SAL regime near the photonic band gap \cite{froufe2017transport} and conjectured that these findings could be generalized to three dimensions
\cite{Edagawa2008,Florescu_PNAS_2009,Cao_PRA_2011,Muller_Scheffold_Adv_Opt_Mat_2014,Froufe_PRL_2016,froufe2017transport}. Early pioneering work by John suggested the possibility of finding SAL in disordered crystalline structures near the band gap \cite{john1996localization,john1987strong,conti2008dynamic}, an idea supported more recently by the numerical studies of Conti and Fratalocci \cite{conti2008dynamic}.  Edagawa and coworkers reported an increase of the inverse participation ratio near the band gap of an amorphous diamond structure, which might indicate the presence of localized states \cite{PhysRevB.82.115116}.  This previous work provides a rationale for the existence of a SAL regime in the vicinity of a band gap of an amorphous photonic material, which we are going to investigate in our work. To this end, we study the transport properties of electromagnetic vector waves in realistic digital representations of three-dimensional hyperuniform silicon networks numerically. We find evidence for the anomalous light transport near the band edge, signaling the onset of Anderson localization at a mobility edge and a broad frequency window where light is localized before the band gap fully develops. 
\section*{Theory} \subsection*{Strong Anderson localization of waves}  The discussion of what defines SAL is very detailed and rich, and for a comprehensive review, we refer to the literature \cite{abrahams1979scaling,zdetsis1985localization,john1996localization,van2000reflection,evers2008anderson,akkermans2007mesoscopic,cherroret2008microscopic,skipetrov2016red}. A working definition for finite-sized systems, proposed by Cherroret and Skipetrov,  is that 'SAL is an interference wave phenomenon in a medium of a finite size that would give rise to truly localized states if the medium were extended to infinity' \cite{cherroret2008microscopic}. The transition to SAL in three dimensions is usually described in the framework of the self-consistent theory (SC) \cite{PhysRevLett.48.699}. It treats localization by introducing a position $\vec{r}$ dependent wave diffusion coefficient $D(\vec{r})$, which decays to zero deep inside the medium \cite{van2000reflection,cherroret2008microscopic}, due to an increased return probability of multiple scattering paths. At the mobility edge, for a semi-infinite medium, one finds the simple algebraic forms  $D(z) = \frac{{D(0)}}{{1 + z/{\xi _c}}}$ and in the localized regime $D\left( z \right) = D \left( 0 \right)\exp \left( { - 2{z \mathord{\left/
 {\vphantom {L \xi }} \right.
 \kern-\nulldelimiterspace} \xi }} \right)$ where $z$ denotes the distance from the surface. The localization length $\xi$ becomes finite at the critical point while $D(z)<D_\text{B}$ already when approaching the mobility edge, where $D_\text{B}=c\ell/3$ denotes the standard Boltzmann diffusion constant. $D(0)$ continues to drop gradually as the localization threshold is crossed \cite{cherroret2010transverse}. For slabs of finite thickness $L$, the transmission coefficient $T(L)$ shows two distinct regimes: initially, it decays diffusive as $\sim L^{-1}$ which is followed by an exponential decay $\sim e^{-L/\xi}$. Precisely at the transition, SC-theory predicts a critical power-law scaling $T\sim L^{-2}$ instead of the exponential decay. 
 
 We point out that the (SC)-theory is an approximate theory, and different variants have been proposed in the literature \cite{sheng2007introduction,cherroret2008microscopic,van2000reflection}. As discussed in ref. \cite{cherroret2008microscopic}, in its original form, it is strictly valid only for $k \cdot \ell \gg 1$. The fact it can describe certain phenomena at the mobility edge and in the localized regime is somewhat fortuitous and not fully understood. However, progress towards a better microscopic understanding has been reported recently \cite{cherroret2008microscopic,tian2008supersymmetric}. Moreover, almost all theoretical and numerical studies were carried out assuming scalar waves and point scatterers randomly distributed in space, i.e., using white-noise Gaussian statistics \cite{skipetrov2018ioffe}. It is not self-evident that the conventional SC-theory \cite{cherroret2010transverse}, can explain the transport of vector waves in spatially correlated, densely filled dielectric materials with a band gap. First of all, the transport $\ell$ and the scattering mean free path $\ell_s$ is not the same for scatterers of size $\sim \lambda$, and the wave is propagating in some effective medium with a wavenumber $k_\text{eff}$ \cite{conley2014light,froufe2017transport}. If and how this affects the predictions by SC-theory is currently not known.  As a consequence of the approximate nature of the SC-theory, in particular, when applied to realistic representations of dielectric materials, we must assume that there is no perfect one-to-one relationship between the macroscopic transport properties, such as the localization length $\xi$ and microscopic quantities like $k=2 \pi/\lambda$ and $\ell$. We, therefore, denote with $(kl)$ the localization parameter, which we assume is similar and proportional but not necessarily identical to $k \cdot \ell$.  For simplicity, we also assume $(kl)_c\equiv 1$ \cite{skipetrov2018ioffe}, and we have tested that using slightly different values for the localization threshold does not significantly affect our findings.
 \newline \indent In summary, in our study the localization parameter $(kl)$ sets the macroscopic properties in the SC-theory, such as $\xi/\ell=6(kl)^2/(1-(kl)^4)$ for $(kl)< 1 $ and $D(0)$, with $D(0)=D_\text{B}(1-(kl)^2)$ for $(kl)\gg 1$\cite{cherroret2010transverse}. The relationship to the microscopic parameter $\ell$ is established via $(kl) \propto k \cdot \ell$ modulus a prefactor of order one that also takes account of the uncertainty with respect to $\ell/\ell_s$ and the effective wavenumber $k_\text{eff}\gtrsim k$. We note that the predictions by SC theory are only meaningful in the limit $\xi/\ell \gg 1$. For example $\xi/\ell > 5$  for $(kl) \in [\frac{3}{4},1)$.
\subsection*{Transport in a photonic band gap }  A photonic crystal with a full photonic band gap (PBG) displays a vanishing density of states (DOS) in the gap. Transport through a finite-sized slab is due to tunneling, characterized by an exponential attenuation with a decay length $L_B$, called the 'Bragg length,' typically on the order of one unit cell, in high refractive index photonic crystals \cite{galisteo2003optical,marichy2016high}. For frequencies outside the gap, or for specific directions in the presence of an incomplete gap, photonic crystals are transparent, before the onset of diffraction \cite{galisteo2003optical}. Perfect crystals show neither photon diffusion nor localization. In his celebrated 1987 paper, John suggested that three-dimensional photonic crystal lattices with moderate disorder may exhibit strong localization of photons \cite{john1987strong}, an idea supported more recently by numerical studies \cite{conti2008dynamic}.
%%%%%%%%%%%%%%%%%%%%%%%%%%%%%%%%%%%%%%%%%%%%%%%%%%%%%%%%%%%%%%%%%%%%%%%%%%%%%%%%%%%%%%%%%%
% Hyperuniform silicon networks with a photonic band gap.
%%%%%%%%%%%%%%%%%%%%%%%%%%%%%%%%%%%%%%%%%%%%%%%%%%%%%%%%%%%%%%%%%%%%%%%%%%%%%%%%%%%%%%%%%%
%\paragraph*{Hyperuniform silicon networks with a photonic band gap.} 
Amorphous photonic materials show similarities but, importantly,  also distinct differences to photonic crystals. Amorphous photonic systems are disordered but the spatial distribution of dielectric material is highly correlated, which can also lead to the opening of a full PBG. The genuine disorder, however, implies that these materials display strong scattering for frequencies outside the gap, or if the gap is incomplete for all frequencies, with the notable exception of transparency in stealthy hyperuniform materials in the long-wavelength limit \cite{Torquato_J_App_Phys_2008,Leseur2016}, which we do not address here. Strong scattering outside the gap opens the possibility for the existence of SAL transport regimes, even in the absence of defect states \cite{froufe2017transport}. 
\section*{Numerical simulations and fitting}
\subsection*{Optical transport simulations and density of states (DOS)} 
\begin{figure}
\includegraphics[width=.95\columnwidth]{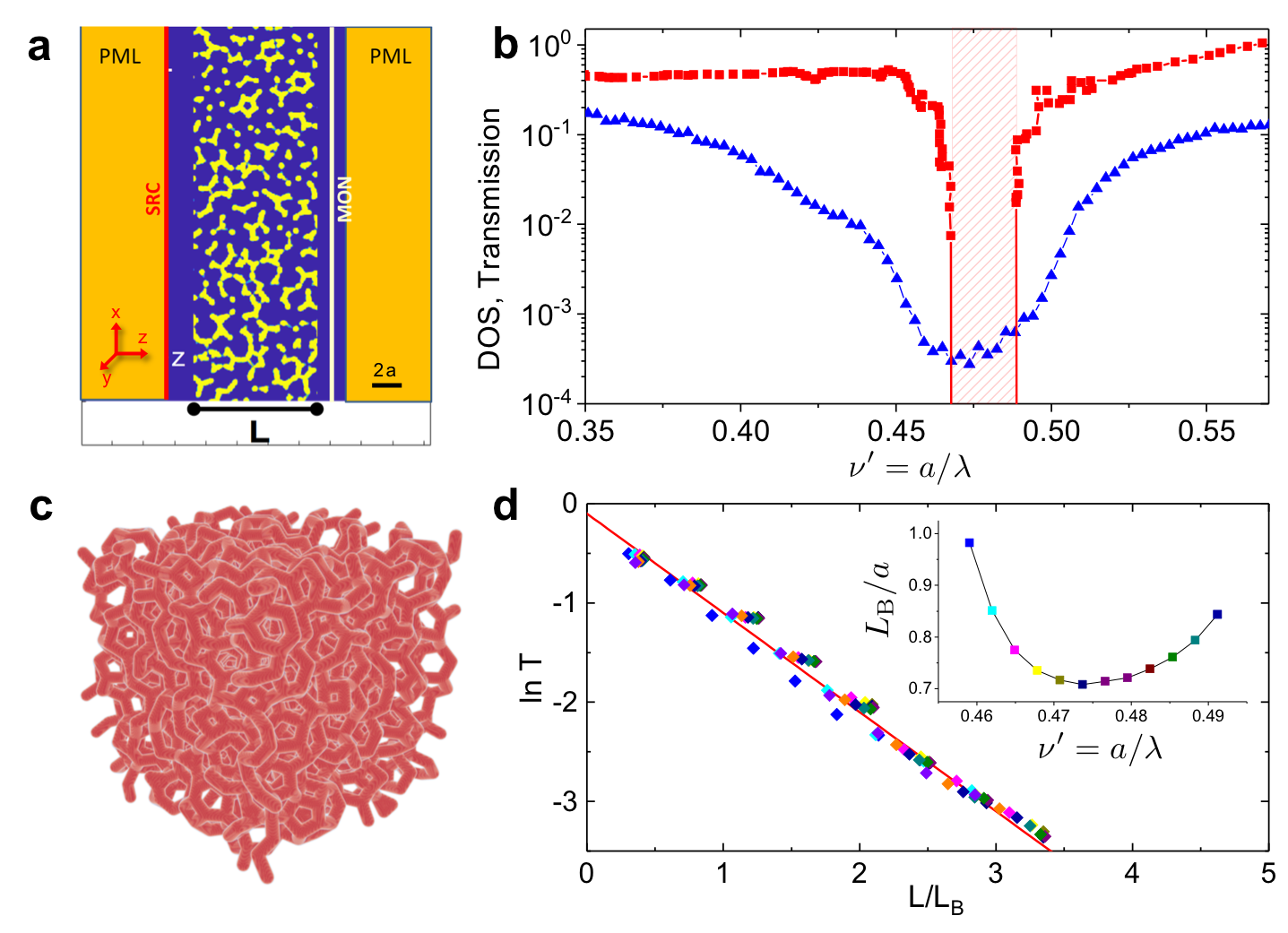}
\caption{\label{fig:1} Numerical simulations of optical transport properties of hyperuniform silicon ($n=3.6$) network structures. {(\bf{a})} Cross section of a three dimensional FDTD (MEEP) simulation box, thickness  $L=6a$. A light wave, linearly polarized along the x-axis, is launched on the left side and propagates along the z-axis. The photonic network structure is terminated with perfectly matched layers (PML) at both sides of the box along the propagation axis. PMLs act as absorbers. The source (SRC) and detector (MON) are placed at a distance approximately $\sim a$ from the sample, which is held in vacuum. Periodic boundary conditions are applied along x and y directions. {(\bf{b})} Triangles: transmittance spectrum $T(a/\lambda)$ for a slab of thickness $L=18a$ for a filling fraction $\phi=0.28$. The optical transport data is compared to numerical calculations of the density of states (DOS) (squares). The band gap center frequency is $\nu^\prime_\text{Gap}=a/\lambda_\text{Gap}=0.478$ and the width $\Delta \nu$ is indicated by the shaded area. {(\bf{c})} Three dimensional rendering of a  hyperuniform network structure, edge length $6 a$ and filling fraction $\phi=0.28$. The size of the structure used in the simulation is $18a \times 18a \times L$ with $L\le 18 a $, which is repeated periodically in (x,y) direction to construct the slab geometry. {(\bf{d})} In the gap the transmittance decays exponentially and $\ln T$ collapses on a master curve when plotted in reduced units $L/L_B$. $L_B\le a$ denotes the Bragg length and it is found to be smallest near the gap center frequency $\nu^\prime_\text{Gap}=0.478$, see also supplemental Figure~\ref{fig:PBG}.} \end{figure}

We study the transport of waves in three-dimensional hyperuniform silicon photonic network structures, refractive index $n=3.6$, derived from the center positions of an assembly of 10'000 randomly close-packed spheres, diameter $a$, as described earlier \cite{Haberko_OPEX_2013,song2008phase}. The design protocol consists of mapping the seed pattern into tetrahedrons by performing a Delaunay tessellation. Then, the centers of mass of the tetrahedrons are connected, resulting in a tetravalent network structure of interconnected rods with the desired structural properties. The diameter of the rods sets the space-filling fraction $\phi$, Figure~\ref{fig:1} {(\bf{a,c})}. The length $a$ is the typical short-range structural length scale of the network, which for a crystal would be the lattice constant.  The seed point pattern is hyperuniform, but not stealthy, meaning that the isotropic structure factor vanishes asymptotically in the limit $S(k)\to 0$ for $k\to 0$ where $k=2 \pi/\lambda$ denotes the wavenumber in vacuum.  Practically identical network structures have been considered by Liew et al. in a study of the optical DOS \cite{Cao_PRA_2011} (see also Supplementary Fig.~\ref{DOS-LIEW}). They report a substantial depletion of the DOS, by more than two orders of magnitude, over a significant range of frequencies, indicating the presence of a band gap for different values of $\phi$. In the present study, we consider networks with a filling fraction of $\phi=0.28$, shown to display the most pronounced photonic properties in their study \cite{Cao_PRA_2011}.
%%%%%%%%%%%%%%%%%%%%%%%%%%%%%%%%%%%%%%%%%%%%%%%%%%%%%%%%%%%%%%%%%%%%%%%%%%%%%%%%%%%%%%%%%%
% Numerical simulations of wave transport.
%%%%%%%%%%%%%%%%%%%%%%%%%%%%%%%%%%%%%%%%%%%%%%%%%%%%%%%%%%%%%%%%%%%%%%%%%%%%%%%%%%%%%%%%%%
\newline \indent For the optical transport simulations, we apply the finite differences time-domain (FDTD) approach, implemented by the MIT Electromagnetic Equation Propagation (MEEP) \cite{oskooi2010meep}. It is considered to be one of the most potent simulation techniques to study electromagnetic wave transport. In a single MEEP-simulation run, a broadband pulse of linearly polarized light, with an electric field vector parallel to one of the sides of the simulation box, is incident on the sample, as illustrated in Figure~\ref{fig:1} {(\bf{a})}. We obtain the full spectrally resolved information about the optical transmittance $T(a/\lambda,L)$, Figure~\ref{fig:1} {(\bf{b})}. We present all spectra in terms of the reduced frequency $\nu^\prime :=a/\lambda=\nu a/c$ where $\lambda$, $\nu$, and $c$ denote the wavelength, frequency and speed of light in vacuum \cite{van2000reflection}. We obtain results for slabs of thickness $L= 0.3 a$ to $18a$ and average the simulations results over 6 (thick slabs) to 15 (thin slabs) independent configurations of the network structure. We note that the networks are structurally isotropic \cite{Haberko_OPEX_2013} and therefore, all incident polarization states on average lead to the same results.  
\newline \indent To obtain the normalized photonic density of states  we use the supercell method \cite{Joannopoulos_book}, implemented in the open-source code MIT Photonic-Bands (MPB),  \cite{Johnson2001_mpb}, Figure~\ref{fig:1} {(\bf{b})}.  Due to computational limitations, we have generated equivalent, but smaller seed patterns with periodic boundary conditions applied. To this end, we are using a packing algorithm developed by Torquato and coworkers~ \cite{skoge2006packing}. 
%%%%%%%%%%%%%%%%%%%%%%%%%%%%%%%%%%%%%%%%%%%%%%%%%%%%%%%%%%%%%%%%%%%%%%%%%%%%%%%%%%%%%%%%%%
%FIGURE 1
%%%%%%%%%%%%%%%%%%%%%%%%%%%%%%%%%%%%%%%%%%%%%%%%%%%%%%%%%%%%%%%%%%%%%%%%%%%%%%%%%%%%%%%%%%
%%%%%%%%%%%%%%%%%%%%%%%%%%%%%%%%%%%%%%%%%%%%%%%%%%%%%%%%%%%%%%%%%%%%%%%%%%%%%%%%%%%%%%%%%%
%FIGURE 1
%%%%%%%%%%%%%%%%%%%%%%%%%%%%%%%%%%%%%%%%%%%%%%%%%%%%%%%%%%%%%%%%%%%%%%%%%%%%%%%%%%%%%%%%%%
% Low order scattering and diffusive transport
%%%%%%%%%%%%%%%%%%%%%%%%%%%%%%%%%%%%%%%%%%%%%%%%%%%%%%%%%%%%%%%%%%%%%%%%%%%%%%%%%%%%%%%%%%
%\paragraph*{Low order scattering and diffusive transport.}  
\subsection*{Fitting procedure} The spectrum of transmittance $T(\nu^\prime,L)$ depends on a number of a priori unknown parameters. Extracting all these parameters from a global fit to SC-theory is not stable and prone to overfitting. To overcome this problem, we first determine parameters that do not scale with $\nu^\prime$, such that eventually, $T(\nu^\prime,L)$ predicted by SC-theory, depends only on the one parameter $(kl)$. Once this is achieved, the predictions by SC-theory and diffusion theory can be compared without fitting bias.
\newline We test the presence of three transport regimes: (i) evanescent transport $\ln[T(\nu^\prime,L)]\sim - L/L_B$ in a PBG with a Bragg length $L_B\sim \ell$.  (ii) diffusive transport with $D(z)\approx \text{const.}$ and (iii) SAL, as described by the SC-theory with $(kl) \lesssim 1$ and $\xi \gg \ell$. We proceed as follows. i) We start by looking for the signature of an evanescent decay in the gap. ii) Next, we consider frequency intervals far below and above the gap, where the predictions for $T(L)$ from classical diffusion theory are sufficient to describe the data. This is the case whenever $L\gg \ell$ and $D(z)\approx \text{const.}$ From this fit we extract the angle averaged reflection coefficient $R$. iii) Next we identify the position(s) of the mobility edge(s) $\nu^\prime_c$, where $\left (kl\right)= \left (kl\right)_c \equiv 1$, by fitting the data with SC-theory treating both $(kl)$ and and $\ell$ as adjustable parameters. From this fit we find the anchor points $\nu^\prime_{c}$ where $\left(kl\right)=1$. Together with $k=2\pi/\lambda_c=2\pi \nu^\prime/a$ this provides us with the proportionality between $k \cdot \ell$ and $(kl)$. iv) Now, that we have fixed all other parameters, we will attempt to describe the entire data set for $T(\nu^\prime,L)$ by SC-theory with only one adjustable parameter $(kl)[\nu^\prime]$. 
\section*{Results}
\subsubsection*{Density of states and tunneling through the gap}
The numerical calculations of the density of states (DOS) reveal a full band gap for frequencies in the interval $\nu^\prime \in [0.47,0.49]$, Fig.~\ref{fig:1} (b). Our results for the gap-center position and the width of the gap are in good agreement with an earlier study by Liew. et al. using a different method but applied to a practically identical system at $\phi=0.28$ volume filling fraction \cite{Cao_PRA_2011}, see also Fig. \ref{DOS-LIEW}. Indeed, for $\nu^\prime \in [0.47,0.49]$, where the DOS is zero, the initial decay of $T(L)$ is exponential with a decay length $L_B< a$, Figure~\ref{fig:1} (d). The decay length rises towards the band edges and is smallest around the center frequency $\nu^\prime_\text{Gap}$. This observation is consistent with tunneling of evanescent waves through the whole sample of thickness $L$ \cite{Joannopoulos_book,marichy2016high}. We find that the evanescent regime appears to extend over a slightly more extensive range of frequencies compared to the band gap. We will address this point again at the end of this section.  
\subsubsection*{Transmission coefficient in the multiple scattering regime} 
\begin{figure}
\centering
\includegraphics[width=.7\columnwidth]{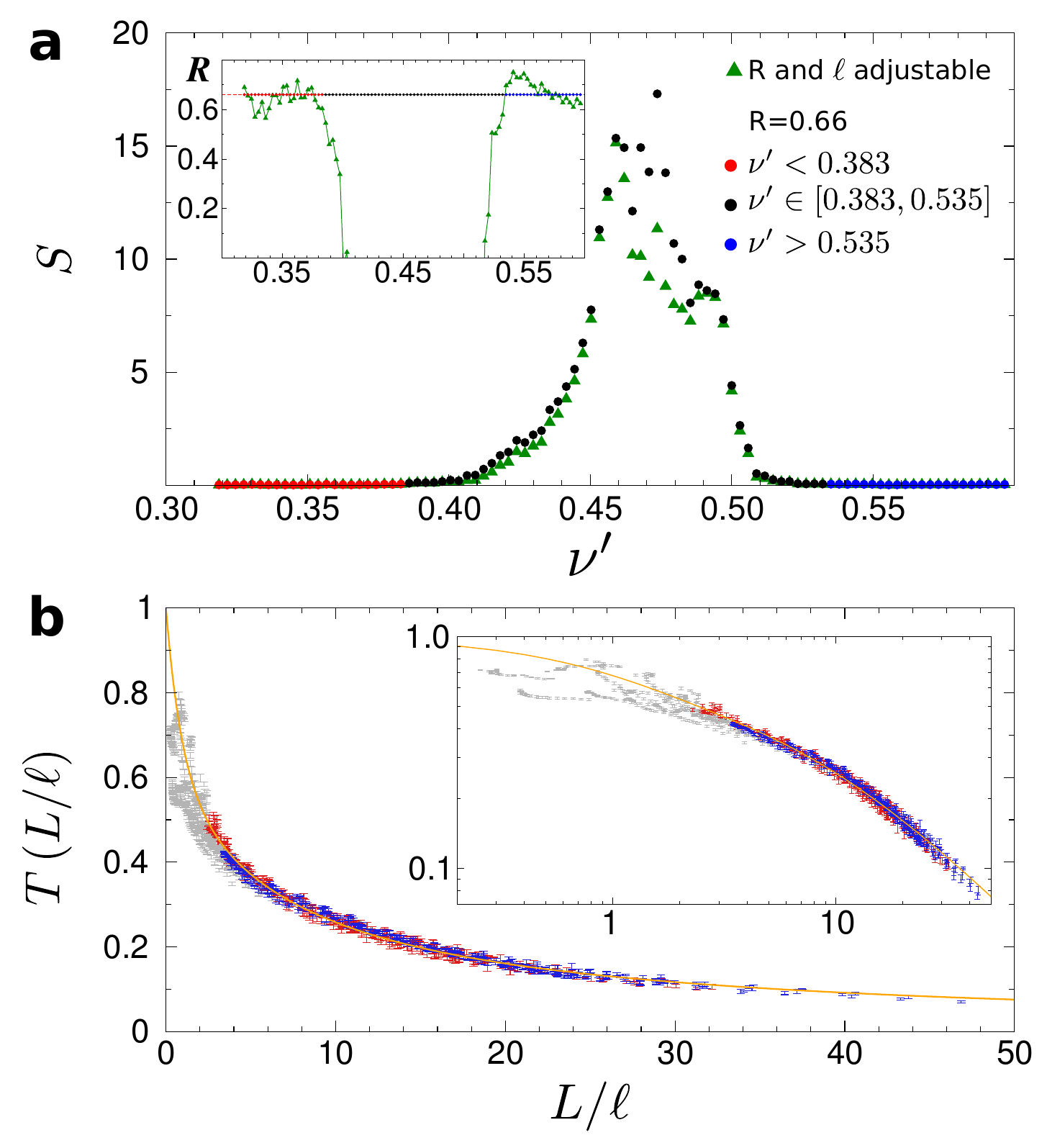}
\caption{\label{fig:2} Comparison to diffusion theory for $(kl)\gg1$. {(\bf{a})} Error of the least squares of the fit of Eq.~\eqref{SCDurian}, in the absence of SAL, to the data with $z_0$ and $\ell$ as adjustable parameters (triangles). In the frequency range marked with red and blue full circles, we find $R=0.66 \pm 0.05$ or $z_0 \simeq 3.25 \pm 0.5$, both in the higher and lower frequency branch (inset). Full circles: results from least squares fitting when setting $R=0.66$ constant. {(\bf{b})} Far from the gap, for $\nu
^\prime \ne [0.383,0.535]$, transport is diffusive and all data for $T(L/\ell)$ can be collapsed on a master curve given by Eq.\eqref{SCDurian} (yellow line) with $\tilde L=L$ and $z_0=3.25$  $(R=0.66)$. Red (blue) full circles refer to the same data at frequencies lower (higher) than $\nu^\prime_\text{Gap}$ shown in {(\bf{a})}. Data for $L<2.7a$, not taken into account for the fitting procedure, are marked with grey symbols.}
\end{figure}

Standard diffusion theory describes transmission through samples whose thickness $L$ is much larger than the mean
free path $\ell$ and thus $T\sim \ell/L \ll 1$. Since the size of our simulation box is limited to $L\le 18 a$, we also have to include data for slabs with thicknesses on the order of a few $\ell$, in particular far from the gap where the transmission coefficient $T$ can be closer to one. To be able to describe the transition from ballistic to diffusive transport as well as to SAL we follow an approach developed by Durian et al. Their theory, which is based on the telegrapher equation, takes into account ballistic transmission, lower order scattering as well as diffusive multiple scattering of light. Their theory accurately predicts $T_{L\to 0}=1$ for thin samples, while diffusion theory fails in this limit. In the methods section we explain how to consistently merge Durian's approach~\cite{lemieux1998diffusing} with the self-consistent theory of localization and we obtain:
\begin{equation}T=\frac{z_{0}+D_B/D\left(0\right)}{2z_{0}+\tilde{L}/\ell}\left(1-e^{-L/\ell}\right)-\frac{\tilde{L}/\ell}{2z_{0}+\tilde{L}/\ell}e^{-L/\ell}+\frac{\eta\left(L/\ell\right)}{2z_{0}+\tilde{L}/\ell}+e^{-L/\ell}\label{SCDurian} \end{equation} 
where $\tilde L $, $D\left(0\right)$ and $\eta$ depend on $\left ( L,\ell\right)$ as well as the localization parameter $\left(k l\right)$. We note that merging Durian's theory with SC-theory is unproblematic. The improvements of the former affect thin slabs $L\sim \ell$ while the latter only affects thick slabs $L>\xi\gg\ell$. In essence, Eq.~\eqref{SCDurian} provides a proper interpolation scheme between the two limiting cases.
In the absence of SAL, for $\left(kl\right)\to\infty$ , we recover Durian's results for $T\left(L/\ell, z_{0}\right)$ with $\tilde L \equiv L$, $D_\text{B}/D(0)= 1$ and $\eta(L/\ell)=0$. Then, for $L\gg \ell$, Eq.~\eqref{SCDurian} reduces to the common expression for diffuse transport  ${T}\left( {L \gg \ell}, z_{0} \right) = \frac{{1 + {z_0}}}{{2{z_0} + L/\ell}}$. \newline When deriving Eq.~\eqref{SCDurian} we did not distinguish between the scattering and the transport mean free path and only use one $\ell$ for the mean free path. By considering an extended version of the model, taking into account the scattering anisotropy parameter $g$, as described in ref. \cite{lemieux1998diffusing}, we have verified that these simplifications do not adversely influence the quality of our fit. Moreover, we neglect specular reflections at the interface of the order of a few percent at most. 
\newline In diffusion theory, the extrapolation length $z_0$ is linked to the angular averaged reflectivity $R$ via $z_{0}= \frac{2}{3}\frac{1+R}{1-R}$.  We determine $R$ directly from a fit to the data using a least-squares fitting procedure. Since the transmittance varies by several orders of magnitude and we are interested in the behavior of $T\left(L\right)$ in different regimes, we choose the natural logarithm of the transmittance as the function to fit. We define
\begin{equation}
S\equiv\sum_{i=1}^N\left[\ln\left(T_{FDTD}\left(L_{i}\right)\right)-\ln\left(T\left(L_{i}/\ell,z_{0}\right)\right)\right]^{2}\label{leastsq}
\end{equation} as the sum of squares to be minimized. The index $i$ runs over the $N=15$~data points from $L/a=2.7$ to $L/a=18$, see table~\ref{tabelVal}. We note that $\ell$ sets the optical thickness in terms of the characteristic length $a$ in our system: $L/a=L/\ell\times\ell/a$. We first fit Eq.~\eqref{SCDurian} to the numerical data, assuming the absence of localization or $\left(k l \right)\to \infty$. Our analysis covers the entire frequency range considered, $\nu^\prime\in [0.3,0.6]$ treating both  $R$ and $\ell$ as adjustable parameters. Figure~\ref{fig:2} (a) shows the frequency dependence of $S$ and the values of $R(\nu^\prime)$ we obtain. Far from the gap center frequency $\left|\nu^\prime-\nu^\prime_\text{Gap}\right| \gtrsim 0.1$ we find excellent agreement between theory and data, as shown in Fig.~\ref{fig:2} (b), indicating a classical transport regime controlled by ballistic transmission, low order scattering which eventually evolves to become diffusive for $L\gg \ell$.  In the same frequency range we find $R=0.66\pm 0.05$ to be approximately constant, corresponding to $z_0 \simeq 3.25 \pm 0.5$. We repeat the fit keeping $R=0.66$ fixed and the goodness of the fit is the same, Figure~~\ref{fig:2}(a). For comparison, we have calculated the internal reflection coefficient from an effective medium interface $n_\text{eff}=1.42$ (for $\phi=0.28$) to vacuum and find a very similar value $R \simeq 0.73$, which demonstrates the consistency of our fitting procedure (for details see Supplementary Fig.~\ref{fig:neff} and~\ref{fig:av_refl}).
%%%%%%%%%%%%%%%%%%%%%%%%%%%%%%%%%%%%%%%%%%%%%%%%%%%%%%%%%%%%%%%%%%%%%%%%%%%%%%%%%%%%%%%%%%
%FIGURE 2
%%%%%%%%%%%%%%%%%%%%%%%%%%%%%%%%%%%%%%%%%%%%%%%%%%%%%%%%%%%%%%%%%%%%%%%%%%%%%%%%%%%%%%%%%%
%%%%%%%%%%%%%%%%%%%%%%%%%%%%%%%%%%%%%%%%%%%%%%%%%%%%%%%%%%%%%%%%%%%%%%%%%%%%%%%%%%%%%%%%%%
%FIGURE 2
%%%%%%%%%%%%%%%%%%%%%%%%%%%%%%%%%%%%%%%%%%%%%%%%%%%%%%%%%%%%%%%%%%%%%%%%%%%%%%%%%%%%%%%%%%

%%%%%%%%%%%%%%%%%%%%%%%%%%%%%%%%%%%%%%%%%%%%%%%%%%%%%%%%%%%%%%%%%%%%%%%%%%%%%%%%%%%%%%%%%%
%Self consistent theory of localization
%%%%%%%%%%%%%%%%%%%%%%%%%%%%%%%%%%%%%%%%%%%%%%%%%%%%%%%%%%%%%%%%%%%%%%%%%%%%%%%%%%%%%%%%%%
%%%%%%%%%%%%%%%%%%%%%%%%%%%%%%%%%%%%%%%%%%%%%%%%%%%%%%%%%%%%%%%%%%%%%%%%%%%%%%%%%%%%%%%%%%
%\paragraph*{Self consistent theory of localization (SC-theory)}  
\subsubsection*{Self consistent theory of localization}  
\begin{figure}
\centering
\includegraphics[width=.6\columnwidth]{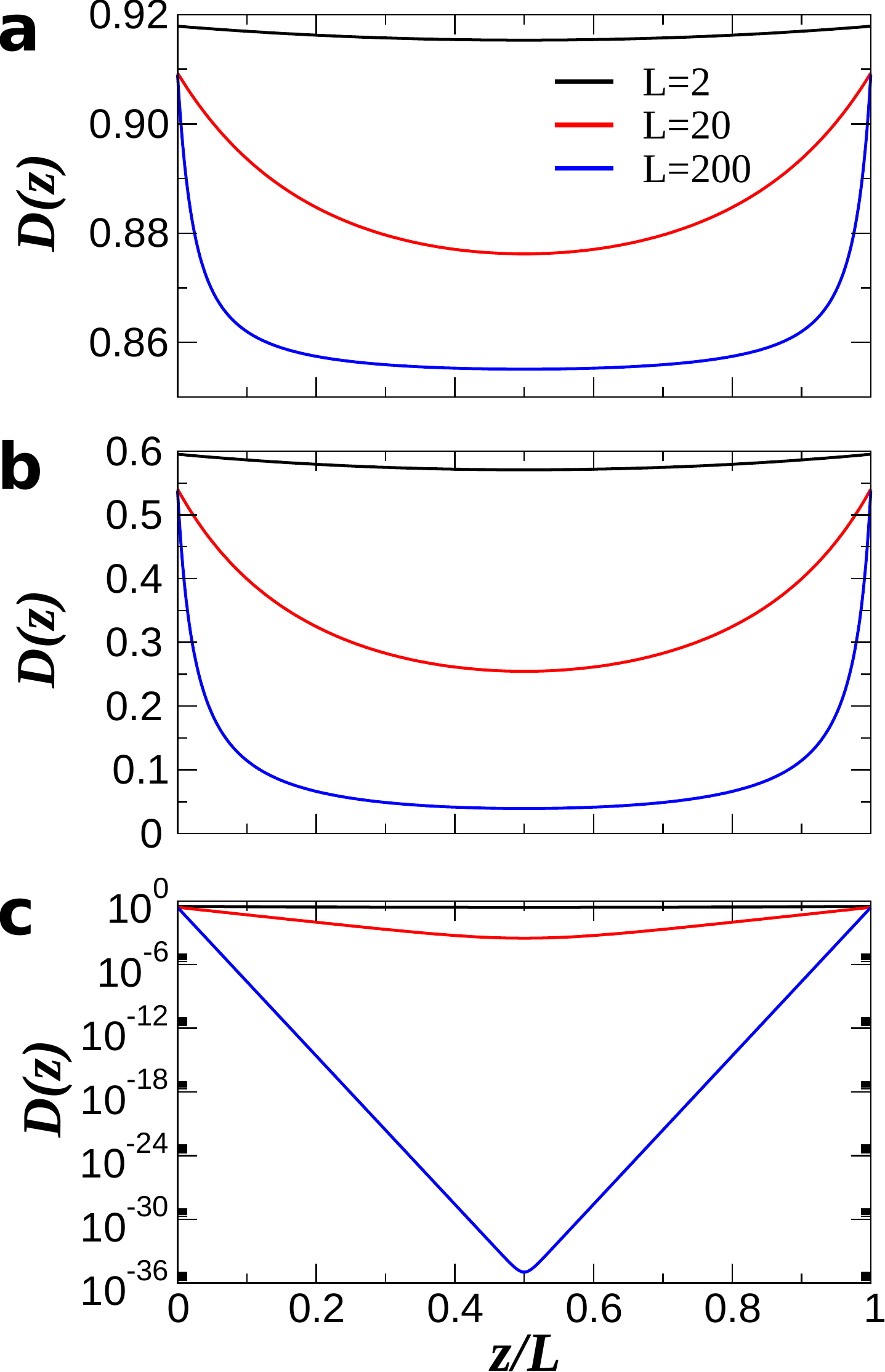}
\caption{\label{fig:3a} Position dependence of the diffusion coefficient in the SAL regime.   $D\left(z/L\right)$ for $\left(k l\right)_{c}=1$, $L/\ell=2$ (black), $L/\ell=20$ (red) and $L/\ell=200$ (blue).{(\bf{a})} $k l=2.6$; {(\bf{b})} $k l=1$; and {(\bf{c})} $k\ell=0.6$
}
\end{figure}

\begin{figure}
\centering
\includegraphics[width=.8\columnwidth]{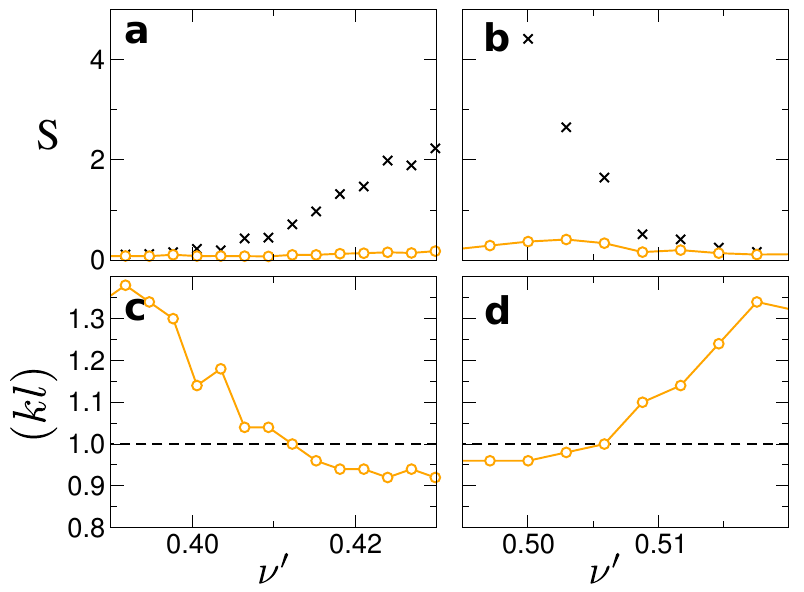}
\caption{\label{fig:3} Two parameter fit to the self consistent theory of localization (SC-theory) with $(kl)$ and $\ell$ as adjustable parameters. Fitting error $S$ for a comparison to diffusion theory (crosses) and SC-theory (open circles) for frequencies below the gap frequency $\nu_\text{Gap}$  {(\bf{a})} and above  {(\bf{b})}. {(\bf{c,d})} Localization parameter $(kl)$ extracted from the fit for the same frequency intervals. We find $(kl)=1$ for  $\nu^\prime_{c,l}=0.412$ ($\ell_{c,l}/a=0.513$), panel {(\bf{c})}, and $\nu^\prime_{c,h}=0.506$ ($\ell_{c,h}/a=0.242$), panel {(\bf{d})}. }
\end{figure}

To assess the breakdown of wave diffusion %and as a critical test for a transition to localization 
near the band edge, we compare our numerical data to the SC-theory of localization.  For slabs of finite thickness $L$ we can calculate $D(z)$ for a given value of $\left( kl\right)$ \cite{cherroret2010transverse}. To illustrate the dependence of the diffusion coefficient on $z$, in Fig.~\ref{fig:3a} we plot the $D$ as a function of $z/L$ for different values of $L/\ell$ and $\left(kl\right)$. As expected, for $\left(kl\right)$ greater than one the diffusion coefficient shows a weak dependence on both the total size of the system and the position. On the contrary, deep in the localization regime, $D(z)$ decays exponentially away from the boundaries. Exactly at the localization transition for $\left(kl\right)=1$, the diffusion coefficient is strongly reduced in the center of the slab although the localization length $\xi$ remains  infinite (for details see Methods).
We integrate $D(z)$ to obtain 
 $\tilde L=\int_0^L \! D_B/D(z) \, \mathrm{d}z$ and the function $\eta\left(L/\ell\right)$. 
 %For $D(0)$ in Eq.~\eqref{SCDurian} we neglect the weak $L-$dependence and use $D(z=0,L/\ell)\simeq D(z=0,L/\ell \to \infty)$. Therefore, the predictions of Eq.~\eqref{SCDurian} only depend on $(kl)$ and $\ell$. In the first fitting-iteration we treat both as adjustable parameters. 
 For $\left(kl\right)$ values larger than one, SC-theory gradually approaches the prediction by diffusion theory and the quality of the fit becomes insensitive to the choice of $\left(kl\right)$. %We thus expect both theories to describe the data equally well. 
 As shown in Figure \ref{fig:3}~(a) and (b) the $S-$values of both diffusion theory and SC-theory become comparable for $\nu^\prime<0.4$ and $\nu^\prime>0.51$, signalling a diffusive transport regime. When approaching the gap from lower (or higher) frequencies, the fit with SC-theory, however, leads to substantially smaller S-values, indicating localization. From the two-parameter fit we find a lower frequency mobility edge $(kl)=1$ at $\nu^\prime_{c,l}=0.412$ ($\ell_{c,l}/a=0.513$) and a higher frequency mobility edge at $\nu^\prime_{c,h}=0.506$ ($\ell_{c,h}/a=0.242$), Figure \ref{fig:3} (c) and (d). Using these anchor-values for the mobility edge, $(kl)$ and $\ell$ are linked via the relation $(kl)=  \ell/\ell_c \times \nu^{\prime}/\nu^\prime_c$. With $(kl)\equiv1$ at the mobility edge, we find $k \cdot \ell=1.33$ for $\nu^\prime_{c,l}=0.412$ and $k \cdot \ell=0.77$ for $\nu^\prime_{c,h}=0.506$. We note that we use these two separate proportionality constants $(1.33,0.77)$ for the comparison between theory and numerical data in the higher and lower frequency branch. In all cases, we perform a least-squares fit to $\ln T$ according to Eq.~\eqref{leastsq}. We have considered other tests, such as $\chi
^2$, but these other tests are often based on assumptions that are probably not met in our case.  It is for example well known that $\ln T$ does not necessarily obey Gaussian statistics in the SAL regime \cite{froufe2017transport}.

%%%%%%%%%%%%%%%%%%%%%%%%%%%%%%%%%%%%%%%%%%%%%%%%%%%%%%%%%%%%%%%%%%%%%%%%%%%%%%%%%%%%%%%%%%
%FIGURE 3
%%%%%%%%%%%%%%%%%%%%%%%%%%%%%%%%%%%%%%%%%%%%%%%%%%%%%%%%%%%%%%%%%%%%%%%%%%%%%%%%%%%%%%%%%%
%%%%%%%%%%%%%%%%%%%%%%%%%%%%%%%%%%%%%%%%%%%%%%%%%%%%%%%%%%%%%%%%%%%%%%%%%%%%%%%%%%%%%%%%%%
%FIGURE 3
%%%%%%%%%%%%%%%%%%%%%%%%%%%%%%%%%%%%%%%%%%%%%%%%%%%%%%%%%%%%%%%%%%%%%%%%%%%%%%%%%%%%%%%%%%

%%%%%%%%%%%%%%%%%%%%%%%%%%%%%%%%%%%%%%%%%%%%%%%%%%%%%%%%%%%%%%%%%%%%%%%%%%%%%%%%%%%%%%%%%%
%FIGURE 4
%%%%%%%%%%%%%%%%%%%%%%%%%%%%%%%%%%%%%%%%%%%%%%%%%%%%%%%%%%%%%%%%%%%%%%%%%%%%%%%%%%%%%%%%%%
\begin{figure}
\centering
\includegraphics[width=.65\columnwidth]{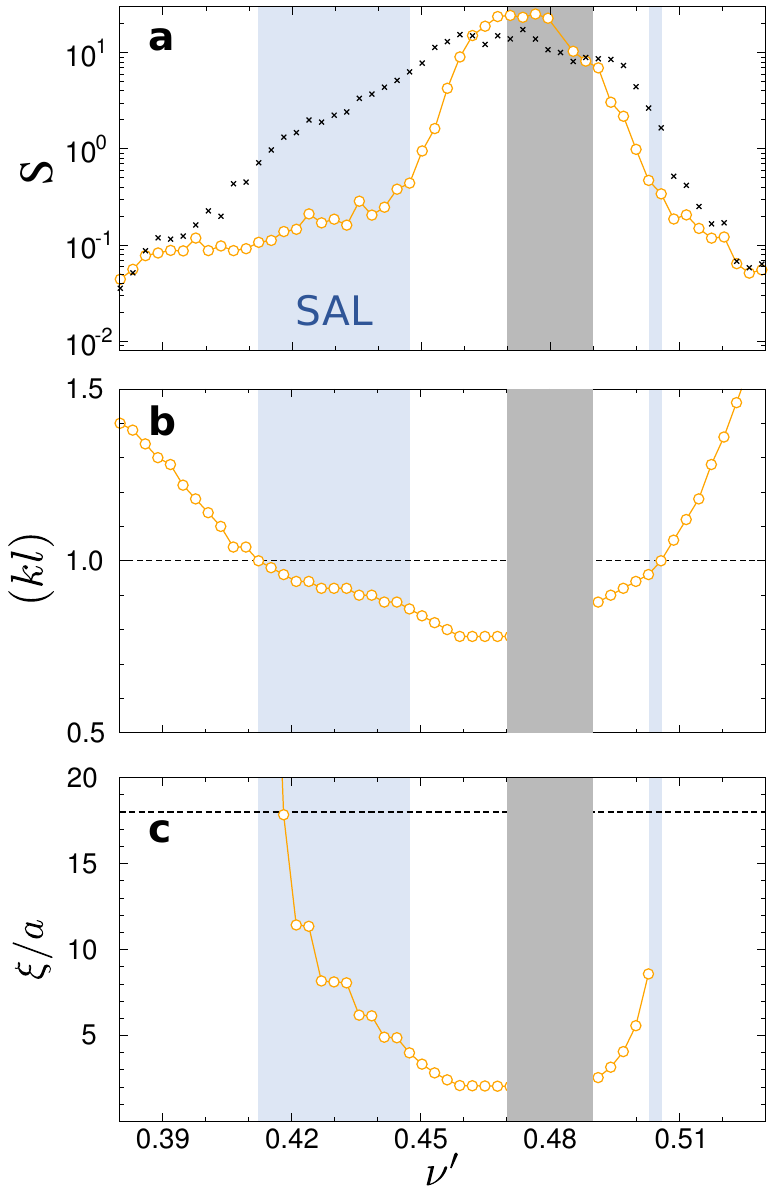}
\caption{\label{fig:4} Single parameter fit to the data using SC-theory (open circles). The only adjustable parameter is $(kl)$ and $\ell$ is defined via $(kl)=  \ell/\ell_c \times \nu^{\prime}/\nu^\prime_c$. Crosses: Fit to the data with diffusion theory, Eq.~\ref{SCDurian} with $\tilde L=L$. {(\bf{a})} Fitting error S (least squares). The blue shaded area indicate where the SC-theory provides an accurate description of the numerical data indicating SAL. {(\bf{b})} Localization parameter extracted from the single-parameter fit. {(\bf{c})} The localization length, given by $\xi/a=6(\ell/a)(kl)^2/(1-(kl)^4)$. The horizontal dashed line indicates the largest accessible length scale in the simulation, given by the edge length of the simulation box $L=18a$.}
\end{figure}
%%%%%%%%%%%%%%%%%%%%%%%%%%%%%%%%%%%%%%%%%%%%%%%%%%%%%%%%%%%%%%%%%%%%%%%%%%%%%%%%%%%%%%%%%%
%FIGURE 4
%%%%%%%%%%%%%%%%%%%%%%%%%%%%%%%%%%%%%%%%%%%%%%%%%%%%%%%%%%%%%%%%%%%%%%%%%%%%%%%%%%%%%%%%%%

%%%%%%%%%%%%%%%%%%%%%%%%%%%%%%%%%%%%%%%%%%%%%%%%%%%%%%%%%%%%%%%%%%%%%%%%%%%%%%%%%%%%%%%%%%
%FIGURE 5
%%%%%%%%%%%%%%%%%%%%%%%%%%%%%%%%%%%%%%%%%%%%%%%%%%%%%%%%%%%%%%%%%%%%%%%%%%%%%%%%%%%%%%%%%%
\begin{figure}
\centering
\includegraphics[width=1\columnwidth]{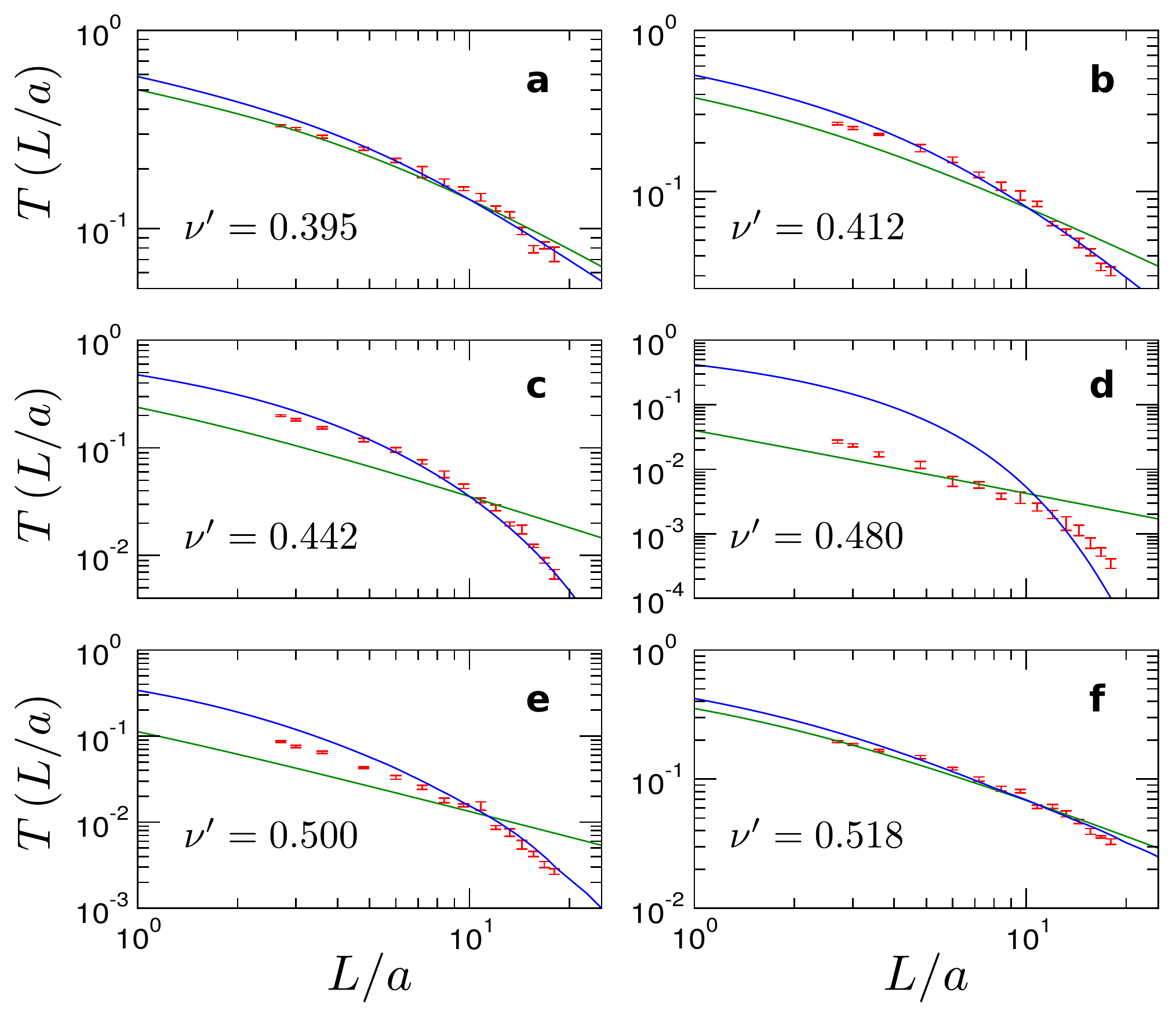}
\caption{\label{fig:5} Transmission coefficient $T$ as a function of the slab thickness $L/a$ in log-log representations for selected frequencies {(\bf{a}-\bf{f})}. Silicon volume fraction $\phi=0.28$.  Only the averaged data for ($L/a\ge2.7$), shown in these plots, were taken into account for fitting. Error bars denote the standard deviation of the results obtained from 15 (thin samples) to 6 (thick samples) realizations. Blue lines: best fit with SC-theory and Eq.
~\eqref{SCDurian}. Green lines: best fit with Eq.~\eqref{SCDurian} for $(kl)\to \infty$, in the absence of SAL.}
\end{figure}
%%%%%%%%%%%%%%%%%%%%%%%%%%%%%%%%%%%%%%%%%%%%%%%%%%%%%%%%%%%%%%%%%%%%%%%%%%%%%%%%%%%%%%%%%%
%FIGURE 5
%%%%%%%%%%%%%%%%%%%%%%%%%%%%%%%%%%%%%%%%%%%%%%%%%%%%%%%%%%%%%%%%%%%%%%%%%%%%%%%%%%%%%%%%%%
Next, we attempt to describe the data with SC-theory over the full range of frequencies $\nu^{\prime}$ using a single adjustable parameter $(kl)$,  as a measure of the distance to the critical point at  $(kl)=1$. We find excellent agreement between SC-theory and the data over the entire range  $\nu^{\prime} \ne [0.45,0.5]$, i.e. outside a central frequency interval in or near the full band gap, Fig.~\ref{fig:4} and Fig.~\ref{fig:5}. The observation of such a single parameter scaling is the key finding of our work. In the regime where $(kl)\lesssim 1.2$ SC-theory describes the data significantly better than diffusion theory and we can describe the data across the critical transition from light diffusion to localization. In the low-frequency branch, between the mobility edge $\nu^\prime_{c,l}=0.412$ and the band edge $\nu^\prime\simeq 0.47$, the sample remains localized over a large frequency interval, comparable or larger in width than the bandgap, as shown in Fig.~\ref{fig:4} (a). In this regime, the localization parameter drops from $(kl)=1$ to about $(kl)=0.85$. In the high-frequency branch, the localized regime is nearly degenerate and the sample rapidly enters the gap after crossing the mobility edge from above for  $\nu^\prime<\nu^\prime_{c,h}$. For $\nu^\prime <0.39$ and $\nu^\prime >0.52$ the position dependence of  $D(z)$ is weak, for our system sizes, and the predictions by SC-theory and by diffusion theory are indistinguishable. 
\newline \indent We would like to add a remark concerning the breakdown of the description of wave transport by SC-theory and the opening of the gap. SC-theory is a heuristic approach that postulates a position-dependent diffusion coefficient $D(z)$ but does not provide a microscopic explanation (which, under certain conditions has been added later \cite{cherroret2008microscopic}). The concept of applying a diffusion equation to describe transport over certain distances $\xi>\ell$ breaks down as $\xi/\ell \to 1$. It it thus unclear whether evanescent decay, Figure \ref{fig:1}{(\bf{d})}, and the correspondingly poor fit with SC-theory, within $\Delta \nu^\prime \simeq 0.01-0.02$ next to the band edge, is due to the breakdown of SC-theory or due other emerging transport phenomena in the vicinity of the gap, as suggested earlier in ref.~\cite{froufe2017transport}. Moreover, we find it interesting to speculate whether the SC-theory could be generalized to provide a unified theory for wave transport in amorphous photonic materials encompassing the transition between diffusive, localized, and band-gap regimes. 
\section*{Summary and conclusions}  In summary, we could show that hyperuniform 3D silicon networks display different characteristic transport regimes for electromagnetic vector waves.
Deep in the gap region, the transmission coefficient decays exponentially, indicating tunneling through the entire sample. Outside the gap region, we observe a critical transition from classical diffusion to  wave localization controlled by a single parameter $(kl)$.  We have shown that in this regime, our numerical data can be described quantitatively by assuming a position-dependent diffusion coefficient $D(z)$ derived from the self-consistent theory of localization (SC). Finding such an agreement is generally understood as evidence for strong Anderson localization of light. 
\section*{Methods}
\subsection*{FDTD Simulations.} 
FDTD Simulations were performed using the MEEP software \cite{oskooi2010meep} and were run on a computer cluster. The network structures were generated using a custom-made code (MATLAB and Statistics Toolbox, The MathWorks, Inc., Massachusetts, United States.) based on the full 10'000 particle seed pattern taken from \cite{song2008phase}. Equivalently, the sicipy.spatial open source library of Python can be used for this purpose. Using a clean cut we obtain slabs of different thickness $L\le18a$ which were then imported to MEEP (see table \ref{tab:lengths} for the exact values of thickness). All units were set to $\mu m$ and the sphere diameter was set to $a=5/3 \mu$m.  We apply periodic boundary conditions in x-and y-directions. Since we do not know the precise size of the simulation box used in ref. \cite{song2008phase}, the periodic boundary conditions in MEEP will not exactly match the original periodic boundary conditions employed when generating the pattern. From our own band structure calculations we find that this can give rise to a few defect states in the gap. We do not expect these defect states to contribute to transport in the diffuse or SAL regime, where the DOS is high, but they can increase the transmittance in the gap for $L\gg L_B$ due to tunneling between defect states. We believe this is the main reason why $T(L)>\exp{(-L/L_B)}$ for $L\gg L_B$ in the gap regime, see e.g. Figure \ref{fig:5} {(\bf{d})}. Since the gap regime is not in the focus of our study, we have not explored this in more detail but plan to address this in future work, using new seed pattern properly matched to the periodic boundary conditions of the MEEP simulation box. 
\begin{table}
\begin{tabular}{|r|r|}
\hline 
$i$ & $L_{i}/a$\tabularnewline
\hline 
\hline 
1 & 2.7\tabularnewline
\hline 
2 & 3.0\tabularnewline
\hline 
3 & 3.6\tabularnewline
\hline 
4 & 4.8\tabularnewline
\hline 
5 & 6.0\tabularnewline
\hline 
6 & 7.2\tabularnewline
\hline 
7 & 8.4\tabularnewline
\hline 
8 & 9.6\tabularnewline
\hline 
9 & 10.8\tabularnewline
\hline 
10 & 12.0\tabularnewline
\hline 
11 & 13.2\tabularnewline
\hline 
12 & 14.4\tabularnewline
\hline 
13 & 15.6\tabularnewline
\hline 
14 & 16.8\tabularnewline
\hline 
15 & 18.0\tabularnewline
\hline 
\end{tabular}
\caption{\label{tab:lengths}Values of $L_{i}/a$ used used
in all fitting procedures, except for the evanescent decay in the band gap.}\label{tabelVal}
\end{table}
\newline The network was illuminated with a broadband pulse of linearly polarized light with electric field vector parallel to one of the sides of the simulation box. The pulse bandwidth was sufficient to cover vacuum wavelengths between 1 $\mu$m and 7 $\mu$m corresponding to reduced frequencies $\nu^\prime \in [0.24,1.67]$. The Poynting vector was recorded on a monitor situated behind the network. Transmittance was calculated by dividing the transmitted power (integral of the Poynting vector over the monitor) by the power transmitted in a reference run (empty simulation box). Perfectly-matched layers (PML) were fitted at both ends of the simulation box and periodic boundary conditions were applied perpenducular to the wave propagation direction. The PML's absorb all transmitted and reflected waves (regardless of incidence direction) and prevent them from re-entering the simulation box due to periodicity. The PML thickness was 7 microns, which ensured that all wavelengths shorter than this value were suppressed. The spatial resolution was equal to 20 pixels per $\mu$m. Convergence tests were performed to check the robustness of the simulation. It was verified that increasing the spatial resolution by a factor of two did not considerably influence the transmittance curves. Also the simulation time was selected in a way to yield robust results. By placing an additional monitor between the source and the network, we checked for flux conservation over the entire frequency interval of interest. 

\subsection*{Numerical implementation of the self consistent theory of localization.}
In the self consistent theory of localization, the standard diffusion
equation is replaced by an equation where the diffusion coefficient
is non-local both in the space and time domain. The renormalization
of the diffusion coefficient accounts for the different return probability
when interference effects are considered in the multiple scattering
regime \cite{cherroret2008microscopic}. We are considering a slab geometry and continuous wave illumination
at a given carrier wave frequency. The simplified geometry, invariant
in the plane parallel to the slab, leads to a set of self-consistent
equations for the diffusion constant dependence and the diffusion
equation Green's function $g\left(z,z^\prime\right)$ in the direction $z$
perpendicular to the slab surfaces (lying between $z=0$ and $z=L$).
Here we work in reduced units, where all lengths are scaled by the mean free path $\ell$. The diffusion
coefficient $D\left(z\right)$ is normalized by the Botzmann diffusion
coefficient $D_\text{B}$. Taking the Fourier transform in the $x,y$ plane,
parallel to the slab boundaries, we reach at the diffusion equation
\begin{subequations} 
\begin{align}
-\frac{\partial}{\partial z}\left[D\left(z\right)\frac{\partial}{\partial z}g\left(q,z,z^\prime\right)\right]+q^{2}D\left(z\right)g\left(q,z,z^\prime\right) & =\delta\left(z-z^\prime\right)\label{eq:SC_10a}\\
g\left(q,z=0,z^\prime\right)-z_{0}D\left(z=0\right)\left.\frac{\partial}{\partial z}g\left(q,z,z^\prime\right)\right|_{z=0} & =0\label{eq:SC_10b}\\
g\left(q,z=L,z^\prime\right)+z_{0}D\left(z=L\right)\left.\frac{\partial}{\partial z}g\left(q,z,z^\prime\right)\right|_{z=L} & =0\label{eq:SC_10c}
\end{align}
\label{eq:SC_10}\end{subequations}
This result together with the self consistent equation for the diffusion coefficient
\begin{equation}
D\left(z\right)=\left[1+\frac{3}{(kl)^2}\int_{0}^{q_{max}^{2}}d\left(q^{2}\right)g\left(q,z,z\right)\right]^{-1}\label{eq:SC_20}
\end{equation}
%(kl)_c
determines the transport properties of our system for all values of
the parameters. In the above equation the cut off is given by $q_{max}=1/3(kl)_{c}^{2}$.
The latter depends on the chosen value $(kl)_c$ with $(kl)=(kl)_c$ at the mobility edge.
\newline \indent We solve the set of self consistent equations by recursively solving
Eq.(\ref{eq:SC_10}) in a first step for all positions of the source
$z^\prime$ and transversal wavenumbers $q$. The solution of the Green
function is then plugged into Eq.(\ref{eq:SC_20}) to correct for
the $z-$dependent diffusion coefficient $D\left(z\right)$ which is inserted back
to Eq.(\ref{eq:SC_10}) until convergence is reached. We choose a discretization scheme where all the considered positions
$z$ and wavenumbers squared $q^{2}$ are evenly spaced. Depending
on the values of the parameters $(kl)$, $(kl)_c$,
and total length $L$, we need to take a step $\Delta z\equiv h$
ranging from $0.02$ up to $0.2$ and between 300 and 600 steps in
$q^{2}$ to achieve a relative precision in $D\left(z\right)$ of
the order or $10^{-4}$ in 5 to 50 recursion steps.
\newline \indent  Taking the second order finite differences approximation for the derivatives in Eq.(\ref{eq:SC_10}) leads to a tridiagonal system of equations which has to be solved for each value of the wavenumber $q$ and
position of the source. Obviously, changing the position of the source
amounts to changing the independent term of the system of equations
and hence all equations for a given value of $q$ can be solved at
once through the inverse of the corresponding tridiagonal matrix.
We choose the Lapack function DGTSV to get the inverse since it is
simple to use and universally accessible while efficient enough for
our purposes \cite{lapack}. 
Specifically, if we take a discretization $z_{i}=h\left(i-1\right)$,
for $i=1,\cdots,n$. And naming $D_{i}\equiv D\left(z=z_{i}\right)$,
$D'_{i}\equiv\frac{dD\left(z=z_{i}\right)}{dz}$, the diagonal terms
of the system of equations read\begin{subequations}
\begin{align}
\mbox{diag}_{1} & =1+\frac{h}{z_{0}D_{1}}\label{eq:SC_30a}\\
\mbox{diag}_{i=2,\cdots,n-1} & =D_{i}\left(2+h^{2}q^{2}\right)\label{eq:SC_30b}\\
\mbox{diag}_{n} & =1+\frac{h}{z_{0}D_{n}}\mbox{,}\label{eq:SC_30c}
\end{align}
\label{eq:SC_30}\end{subequations}the subdiagonal terms are \begin{subequations}
\begin{align}
\mbox{subdiag}_{i=1,\cdots,n-2} & =\frac{D'_{i+1}h}{2}-D_{i+1}\label{eq:SC_40a}\\
\mbox{subdiag}_{n-1} & =-1.\label{eq:SC_40b}
\end{align}
\label{eq:SC_40}\end{subequations}Analogously, the superdiagonal
terms are\begin{subequations}
\begin{align}
\mbox{superdiag}_{1} & =-1\label{eq:SC_50a}\\
\mbox{superdiag}_{i=2,\cdots,n-1} & =-\frac{D'_{i}h}{2}-D_{i}\mbox{.}\label{eq:50b}
\end{align}
\label{eq:SC_50}\end{subequations}Since the sources are located in
the interior of the slab, the points at which the source is located
are $z^\prime_{i=2,\cdots,n-1}=h\left(i-1\right)$. The independent term
vector $T_{i}$ for the source at $z_{i}$ is $\left(T_{i}\right)_{j}=h\delta_{i,j}$. 
\newline Once the numerical solution is obtained for all the source positions,
the value of $g\left(z_{i},z_{i}\right)$ is available for $i=2,\cdots,n-1$.
The $g\left(z_{1},z_{1}\right)$ and $g\left(z_{n},z_{n}\right)$
are obtained by second order accuracy extrapolation of the solutions
$g\left(z_{i},z_{i}\right)$ to the boundary. The full solution is
then used to calculate the integral in Eq.(\ref{eq:SC_20}) by $3/8$
Simpson's rule. The updated value of the diffusion coefficient $D\left(z\right)$
is used in the next step of recursion. The algorithm stops when the
logarithm of an update of $D\left(z\right)$ differs by less than
$10^{-4}$ from the previous value at all the points in the discretization.
The seed $D\left(z\right)$ used to start the recursions is set to
$1$. In all cases $(kl)_c=1$. 

\subsection*{Transmission through a slab using self-consistent theory of localization and the scattering and transport theory for thin slabs.}
In the following we describe how to consistently merge SC-theory with the theory by Durian and coworkers~\cite{lemieux1998diffusing}. %We combine the previous results of SC theory (see supplementary material for details) with the two-stream theory, including ballistic transmission \cite{durian1996two}, in order to get an expression for the full transmission through a slab (under normal incidence, homogeneous and continuous illumination). 
The contribution that is affected by SC-theory is the total diffusive transmission probability $T_d$ (see also \cite{durian1996two}). We compute it considering that the sources of diffuse intensity are continuously  distributed across the slab with an intensity proportional to the ballistic intensity $I_{b}=\exp\left(-z\right)$. To simplify the notation, in this section, we work in reduced units where all lengths are normalized to $\ell$ and $D(z)$ is normalized to $D_B$. The transmittance, neglecting boundary reflectivity of a few percent, is hence 
% \begin{equation}
%T_{d}=\frac{1}{1-F^{2}}\int_{0}^{L/\ell}\left[T_{d}\left(z\right)+F\l%eft(1-T_{d}\left(z\right)\right)\right]e^{-z}dz\textrm{,}\label{eq:du%_10}
%\end{equation}
%where $F\equiv R_{b}e^{-L/\ell}$, and $R_{b}$ being the normal incidence ballistic reflection coefficient at one interface 
%\begin{equation}
%R_{b}=\left(\frac{1-n_{eff}}{1+n_{eff}}\right)^{2}\textrm{.}\label{eq%:du_20}
%\end{equation}
%We have shown that $n_{eff}\simeq1.417$, that leads to $R_{b}\simeq0.03$.
%In the following we shall neglect the terms $F$. We have
\begin{equation}
T_{d}=\int_{0}^{L}T_{SCT}\left(z\right)e^{-z}dz\label{eq:du_30}
\end{equation}
where $T_{SCT}\left(z\right)$ is the diffuse transmission probability for a source at $z=z^\prime$
(see Supplementary Material, Eqs. \eqref{eq:tr_10}-\eqref{eq:lvzb}, for details), explicitly we have
\begin{equation}
T_{d}=\int_{0}^{L}\frac{z_{0}+\int_{0}^{z}\frac{1}{D\left(x\right)}dx}{2z_{0}+\tilde{L}}e^{-z}dz\label{eq:du_40}
\end{equation}
with
\begin{equation}
\tilde{L}\equiv\int_{0}^{L}\frac{1}{D\left(x\right)}dx\label{eq:tr_130}
\end{equation}
The first term ($z_{0}$ in the numerator) can be explicitly integrated:
\begin{equation}
\int_{0}^{L}\frac{z_{0}}{2z_{0}+\tilde{L}}e^{-z}dz=\frac{z_{0}}{2z_{0}+\tilde{L}}\left(1-e^{-L}\right)\label{eq:du_50}
\end{equation}

The second term in Eq.(\ref{eq:du_40}) is more involved since $D\left(z\right)$
is not known a priori, the integral to be solved is
\begin{equation}
I\equiv\frac{1}{2z_{0}+\tilde{L}}\int_{0}^{L}e^{-z}\int_{0}^{z}\frac{1}{D\left(x\right)}dxdz\label{eq:du_60}
\end{equation}
that can be formally integrated by parts using $u\left(z\right)\equiv\int_{0}^{z}\frac{1}{D\left(x\right)}dx$
and $dv\left(z\right)\equiv e^{-z}dz$. Hence
\begin{equation}
I=\frac{1}{2z_{0}+\tilde{L}}\left[-e^{-L/\ell}\tilde{L}+\int_{0}^{L}dz\frac{e^{-z}}{D\left(z\right)}\right]\label{eq:du_70}
\end{equation}

The second term in the rhs of Eq.(\ref{eq:du_70}) can again be formally
integrated by parts using $u\left(z\right)\equiv1/D(z)$, $dv\left(z\right)\equiv e^{-z}dz$
to give
\begin{equation}
\int_{0}^{L}dz\frac{e^{-z}}{D\left(z\right)}=\frac{1}{D\left(0\right)}\left(1-e^{-L}\right)+\int_{0}^{L}\frac{e^{-z}}{D\left(z\right)}\frac{d\ln\left[D\left(z\right)\right]}{dz}dz\label{eq:du_80}
\end{equation}

Collecting results, we have
\begin{equation}
T_{d}=\frac{z_{0}+1/D\left(0\right)}{2z_{0}+\tilde{L}}\left(1-e^{-L}\right)-\frac{\tilde{L}}{2z_{0}+\tilde{L}}e^{-L}+\frac{\eta\left(L\right)}{2z_{0}+\tilde{L}}\textrm{,}\label{eq:du_90}
\end{equation}
where
\begin{equation}
\eta\left(L\right)\equiv\int_{0}^{L}\frac{LN\left(z\right)}{D\left(z\right)}e^{-z}dz\textrm{ , and }LN\left(z\right)\equiv\frac{d\left[\ln\left(D(z)\right)\right]}{dz}\label{eq:du_100}
\end{equation}
Finally, adding the ballistic transmission probability $e^{-L}$, we obtain  Eq.\eqref{SCDurian} for the total transmision coefficient (transmittance)
\begin{equation}
T\left(L\right)=\frac{z_{0}+1/D\left(0\right)}{2z_{0}+\tilde{L}}\left(1-e^{-L}\right)-\frac{\tilde{L}}{2z_{0}+\tilde{L}}e^{-L}+\frac{\eta\left(L\right)}{2z_{0}+\tilde{L}}+e^{-L}\label{eq:du_120}
\end{equation}
\newline It is worth noticing at this point that in the limit of standard diffusion theory,
$D\left(z\right)=1$ (i.e. the diffusion coefficients coincides with
the standard $D_{B}=c\ell/3$) and $\eta\left(L\right)=0$,since $LN\left(z\right)=0$ 
%according to the definition
%\begin{equation}
%L\left(z\right)\equiv\frac{d\left[\ln\left( D(z)\right)\right]}{dz}
%\end{equation}
%, we have \tilde{L}=L$ and \begin{equation}
%T_{d}=\frac{z_{0}+1}{2z_{0}+L}\left(1-e^{-L}\right)-Le^{-L}\label{eq:%du_110}
%\end{equation}
%In the main text we used the exact same equation (\ref{eq:du_120}) explicitly expressing $\ell$
%and $D_B$, Eq.\ref{SCDurian}, which we had omitted here for brevity.
%non-normalized lengths or diffusion constant, in summary we have\begin{subequations}
%\begin{align}
%T_{SC}\left(L/\ell\right) & =\frac{z_{0}+D_{B}/D\left(0\right)}{2z_{0}+\tilde{L}/\ell}\left(1-e^{-L/\ell}\right)-\frac{\tilde{L}/\ell}{2z_{0}+\tilde{L}/\ell}e^{-L/\ell}+\frac{\eta\left(L\right)}{2z_{0}+\tilde{L}/\ell}+e^{-L/\ell}\label{eq:du_130_1}\\
%\tilde{L} & =\int_{0}^{L}\frac{D_{B}}{D\left(z\right)}dz\label{eq:du_130_b}\\
%\eta\left(L\right) & =\int_{0}^{L}e^{-z/\ell}\frac{D_{B}}{D\left(z\right)}\frac{d\ln\left[D\left(z\right)/D_{B}\right]}{dz}dz\label{eq:du_130_c}
%\end{align}\label{eq:du_130}\end{subequations}

%\bibliographystyle{abbrv}

\section*{Data availability}
All numerical data displayed in the manuscript for $T(L/a)$ as obtained from FDTD calculations and from the self-consistent theory of localization (SC-theory) will be uploaded to the repository Zenodo (\url{https://zenodo.org}). Additional data sets generated and/or analysed during the current study can be generated from the codes uploaded to the repository, or can be obtained upon reasonable request. 
\section*{Code availability}
The codes used to produce the results of this study are included in the repository and described in the main text or the supplementary material. With respect to third party codes, such as the open source codes MPB and MEEP, we refer to the original publications, see refs~\cite{oskooi2010meep,Johnson2001_mpb,skoge2006packing}.

%All additional the data sets generated during and/or analysed during the current study are available from the corresponding author on reasonable request.

%\section*{Code availability}
%The codes used to produce the results of this study are based on a proprietary software library as documented in detail in the manuscript and supplementary material. With respect to third party codes we refer to the original authors, see ref  \cite{oskooi2010meep} and \cite{skoge2006packing}.

\section*{Competing interest}
The authors declare no competing interest.

\section*{AuthorContributions}
F.S. and J.H. designed the study.  J-H. carried out the FDTD calculations using MEEP. LFP did the band structure calculations using MPB. LFP did all the fitting and comparison to SC-theory. FS wrote the paper with contributions by all authors. 

\section*{Acknowledgments}
 We thanks Seng Fatt Liew and Hui Cao for discussions and for sharing the DOS data published in \cite{Cao_PRA_2011}. We thank Sergey Skipetrov for discussion and for providing us with the original FORTRAN code to numerically solve the SC-theory equations, which we have implemented and adapted as described in the text and Supplementary material. F.S. thanks Dave Pine for discussion and comments. This research was supported by the Swiss National Science Foundation through the National Centre of Competence in Research \emph{Bio-Inspired Materials} and through project number, 169074, 188494 (FS) and 192340 (LFP). JH acknowledges the use of the TeraACMIN computer cluster at the Academic Centre for Materials and Nanotechnology, AGH-UST, Krakow, Poland.

%\section*{Data} The numerical data presented in the manuscript for T(L) obtained from FDTD calculations and from the self-consistent theory of localization will be uploaded to a repository.

%\section*{Competing interest}  The authors declare no competing interest. 

%\section*{Methods}

%Here you should list the contents of your Supplementary materials -- below is an example. 
%You should include a list of Supplementary figures, Tables, and any references that appear only in the SM. 
%Note that the reference numbering continues from the main text to the SM.
% In the example below, Refs. 4-10 were cited only in the SM.  
\clearpage
\setcounter{equation}{0}
\setcounter{figure}{0}
\setcounter{table}{0}
\setcounter{page}{1}
\renewcommand{\theequation}{S\arabic{equation}}
\renewcommand{\thefigure}{S\arabic{figure}}
\section*{JH et. al.} 
\section*{Supplementary Material}
\begin{figure}
 \centering
\includegraphics[width=.75\columnwidth]{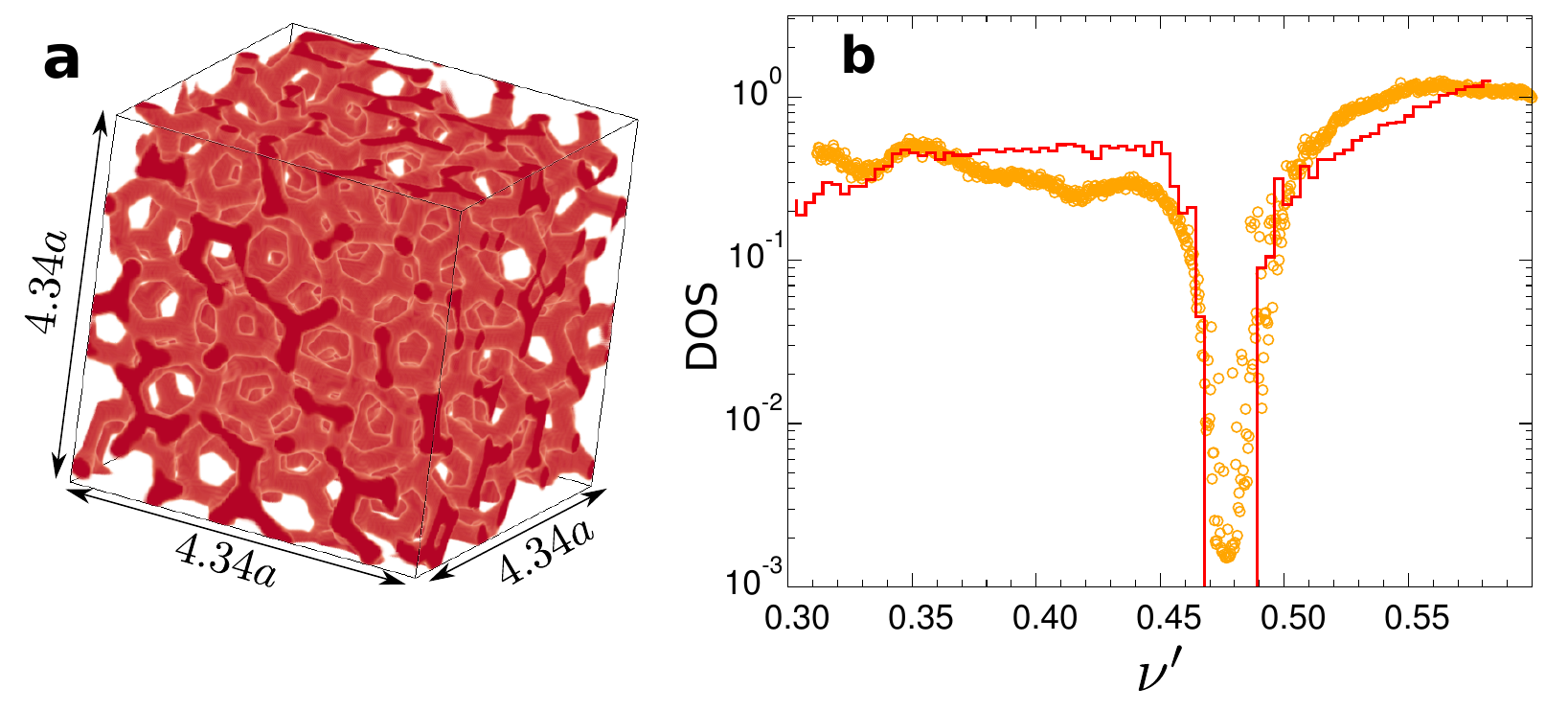}
\caption{\label{DOS-LIEW} {(\bf{a})} Network with $n=3.6$ used to calculate
the band structure with the MPB software. {(\bf{b})} DOS
obtained by Liew et.al. \cite{Cao_PRA_2011} (open symbols) compared
with a frequency histogram of our calculated band structure (solid
line). We note that in ref.\cite{Cao_PRA_2011} $a$ denotes the mean spacing between
spheres of radius $R$ while here we plot both data sets with $a=2R$, where $R$ is the diameter of
the sphere. Our value of $a$ is smaller by a factor $0.9348$ compared to the value used in \cite{Cao_PRA_2011}.}
\end{figure}

%\subsection*{Methods}
\paragraph*{Density of states and comparison to results by Liew  et al.} 
Practically identical network structures have been studied by Liew et al. in a recent study of the optical DOS \cite{Cao_PRA_2011}.  They report a strong depletion of the DOS, by more than two orders of magnitude, over a significant range of frequencies, indicating the presence of a band gap. Moreover, they show that networks with a filling fraction of $\phi=0.28$, display the most pronounced photonic properties \cite{Cao_PRA_2011}. In Figure~\ref{DOS-LIEW}b we compare our band structure calculations, using the supercell method implemented in MPB \cite{Johnson2001_mpb}, to their results using a different algorithm. We are not familiar with the details of their numerical method but apparently there are  limitations in accuracy or convergence which lead to a finite density of states even in the gap center. Our results, using the established supercell method, with periodic boundary conditions, confirms the existence of a full band gap in our system \cite{Florescu_PNAS_2009,Froufe_PRL_2016,Johnson2001_mpb}. More specifically, we start by considering a cubic lattice of lattice
parameter $L$. In order to avoid the emergence of surface states
due to the non conformity of the periodic cell with the actual structure,
we consider a lattice basis which is precisely a cube of length $L$
generated with periodic boundary conditions.

We use the MPB software \cite{Johnson2001_mpb} to obtain the eigenfrequencies
of the photonic crystal. On the one hand, if a stop band opens between
bands $N$ and $N+1$, we expect $N\propto\left(L/a\right)^{3}$,
i.e. we expect $N$ to be proportional to the number of scattering
units in the primitive cell. On the other hand, for a given spatial
resolution $\delta x$, the description of the permittivity contains
$\mathcal{O}\left(\frac{L}{\delta x}\right)^{3}$ voxels in the cell,
and we need $\mathcal{O}\left(\frac{L}{a}\right)^{3}$ bands as discussed
above. It is clear that a calculation with high resolution and large
supercells will be prohibitively expensive both in terms of memory
and computer time. In particular, our model network derived from the
pattern used in the main text \cite{song2008phase} does not fit into
our available computational facilities. However, a smaller system
with $L\simeq4.34a$ and a resolution of $128^{3}$ voxels fits into
the memory and can be solved for 8 different points in the reciprocal
space with reasonable computational efforts and a resolution being
comparable to the one used in transport calculations. In this case
the computation of the first thousand eigenfrequencies for each k
point is enough to cover the spectral region of interest.

In order to generate a suitable seed pattern, we obtain a random closed
packing of $N_{S}$ rigid spheres in a periodic cube. We used the
software developed by Skoge and co-workers \cite{skoge2006packing}.
Setting a packing fraction $\phi_{s}\simeq64\%$, we obtain a random close packed distribution. The centers of the spheres
are the seed pattern to generate the random four-valent network as explained
in the main text. Setting the appropriate diameter of cylinders linking
the nodes of the structure, we obtain a perfectly periodic structure
with filling fraction $\phi=28.14\%$ which is, with high precision,
the same as the one used for transport calculations considered in
the main text. We notice at this point that the Delaunay tessellation
is performed in a $3\times3\times3$ replica of the original seed
pattern. Cutting the final structure to the original size of the cube
avoids any spurious surface effect in the final network.

In Figure \ref{DOS-LIEW} we summarize our results for one of
the structures. In panel a, we represent the spatial structure of
the permittivity as interpreted by the MPB software. Interestingly, using
a seed pattern with only a hundred spheres is enough to generate a
relatively large network. Despite the small number of points in the
seed pattern we do not find obvious spurious patterns in the structure.
In panel b, we compare the DOS as presented by Liew et al. \cite{Cao_PRA_2011} together with a histogram of all the frequencies given
by the MPB software of the structure depicted in panel a (all other
parameters kept as in the main text). The dip in the DOS corresponds almost exactly with
the exact zero of the frequency histogram (both data sets vertically
scaled to occupy the same range). In the MPB we sampled the k space
using 8 points distributed along the $\Gamma-R-X-M-\Gamma$ path in
the reciprocal space with one interpolating point between the vertices.

\begin{figure}[ht]
 \centering
\includegraphics[width=0.7\columnwidth]{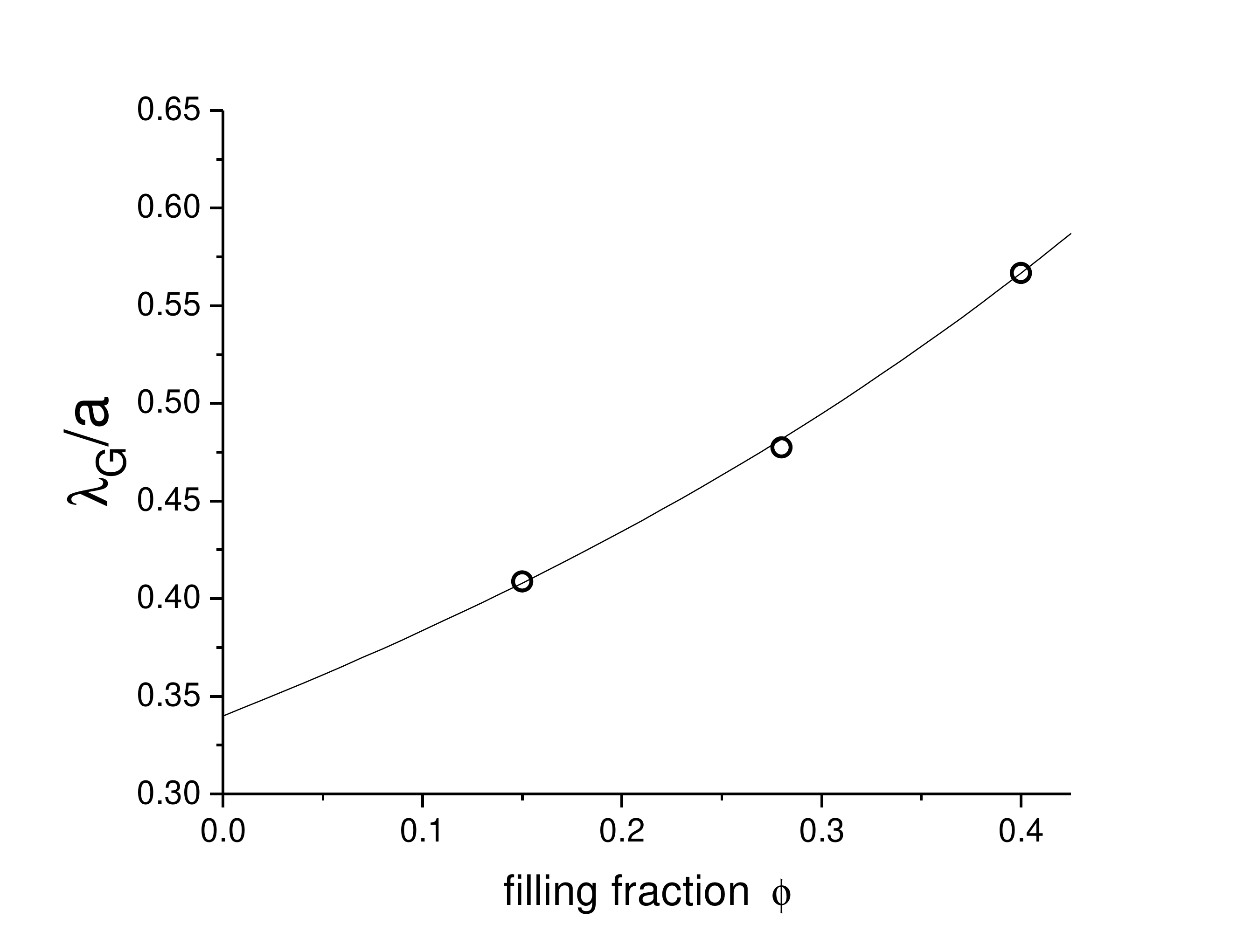}
\caption{\label{fig:neff} Open symbols: Band gap center frequency $\nu^\prime_\text{Gap}(\phi)=a/\lambda_\text{Gap}(\phi)$ extraced from ref.~\cite{Cao_PRA_2011} for different silicon filling fractions $\phi$. Line: Best fit to
$\nu^\prime_\text{Gap}(\phi)=\nu^\prime_\text{Gap}(0) \times n_\text{eff}(\phi)$ with $n_\text{eff}(\phi)$ using the unified mixing formula, Eq.~\ref{mixingform}. Parameters are $\alpha=0.7$ with $\nu^\prime_\text{Gap}(0)=a/\lambda_\text{Gap}(0)=0.34$. For $\phi=0.28$ we find $n_\text{eff}=1.42$.}
\end{figure}

\paragraph*{Effective refractive index.} 
As shown in \cite{Cao_PRA_2011}, data reproduced in Figure \ref{fig:neff}, the gap is red-shifted when increasing the volume fraction of the higher refractive index component. In the long wavelength limit $\lambda \gg a$, i.e. for wavelengths large compared to the size of dielectric heterogeneities, the well known Maxwell-Garnet mixing formula can be applied. If, however, heterogeneities are present on length scales comparable to the wavelength, the situation becomes more complicated. %A detailed discussion of this question is beyond the scope of this work.
Different mixing formulas have been proposed to determine an effective permittivity or effective refractive index. A set of commonly used formulas can be written in a unified way, as follows (for details we refer to the textbook by Sihvola \cite{sihvola1999electromagnetic}): 
\begin{equation}\label{mixingform}\frac{{{\varepsilon _\text{eff}} - {\varepsilon _e}}}{{{\varepsilon _\text{eff}} + 2{\varepsilon _e} + \alpha \left( {{\varepsilon _\text{eff}} - {\varepsilon _e}} \right)}} = \frac{{\phi \left( {\varepsilon  - {\varepsilon _e}} \right)}}{{\varepsilon  + 2{\varepsilon _e} + \alpha \left( {{\varepsilon _\text{eff}} - {\varepsilon _e}} \right)}}\end{equation}
$\phi$ denotes the filling fraction, $\varepsilon$ and $\varepsilon _e$ are the dielectric permittivity of the dispersed or structured material and the background environment, respectively, and  $n_\text{eff}=\sqrt{\varepsilon _\text{eff}}$ is the effective refractive index. For our case $\varepsilon=3.6^2=12.96$ for silicon and $\varepsilon _e=1$ for air. For a given value of $\alpha$ %has to be solved numerically to obtain
we can calculate $\varepsilon _\text{eff}$ from Eq.\eqref{mixingform} and thus obtain $n_\text{eff}$ for all possible compositions $\phi$. Depending on the choice of the dimensionless parameter $\alpha$ different mixing rules are recovered: ($\alpha=0$) Maxwell-Garnett, ($\alpha=1$) Polder-van Santen or Bruggemann and ($\alpha=2$) Coherent Potential approximation. 
Here we treat $\alpha$ as an adjustable parameter to obtain a best fit to the $\nu^\prime_\text{Gap}(\phi)=a/\lambda_\text{Gap}(\phi)$, with $\nu^\prime_\text{Gap}(\phi)=\nu^\prime_\text{Gap}(0) \times n_\text{eff}(\phi)$, Figure~\ref{fig:neff}. We find $\alpha \simeq 0.7$ and $\nu^\prime_\text{Gap}(0)=0.34$. Thus for $\phi=0.28$ we find $n_\text{eff}\simeq 1.42$.
\paragraph*{Angle Averaged Reflection} 
An important parameter in both standard diffusion theory and self
consistent theory of localization is the extrapolation length $z_{0}$.
If internal reflections are not considered, $z_{0}\ell=2\ell/3$ (we
set $\ell^{*}=\ell_{s}\equiv \ell$). Considering the effect of internal reflection
from the effective medium to vacuum leads to a larger $z_{0}$ that
can be written as 
\begin{equation}
z_{0}=\frac{2}{3}\frac{1+R}{1-R}\textrm{,}\label{eq:R_10}
\end{equation}
\begin{figure}
 \centering
\includegraphics[width=.5\columnwidth]{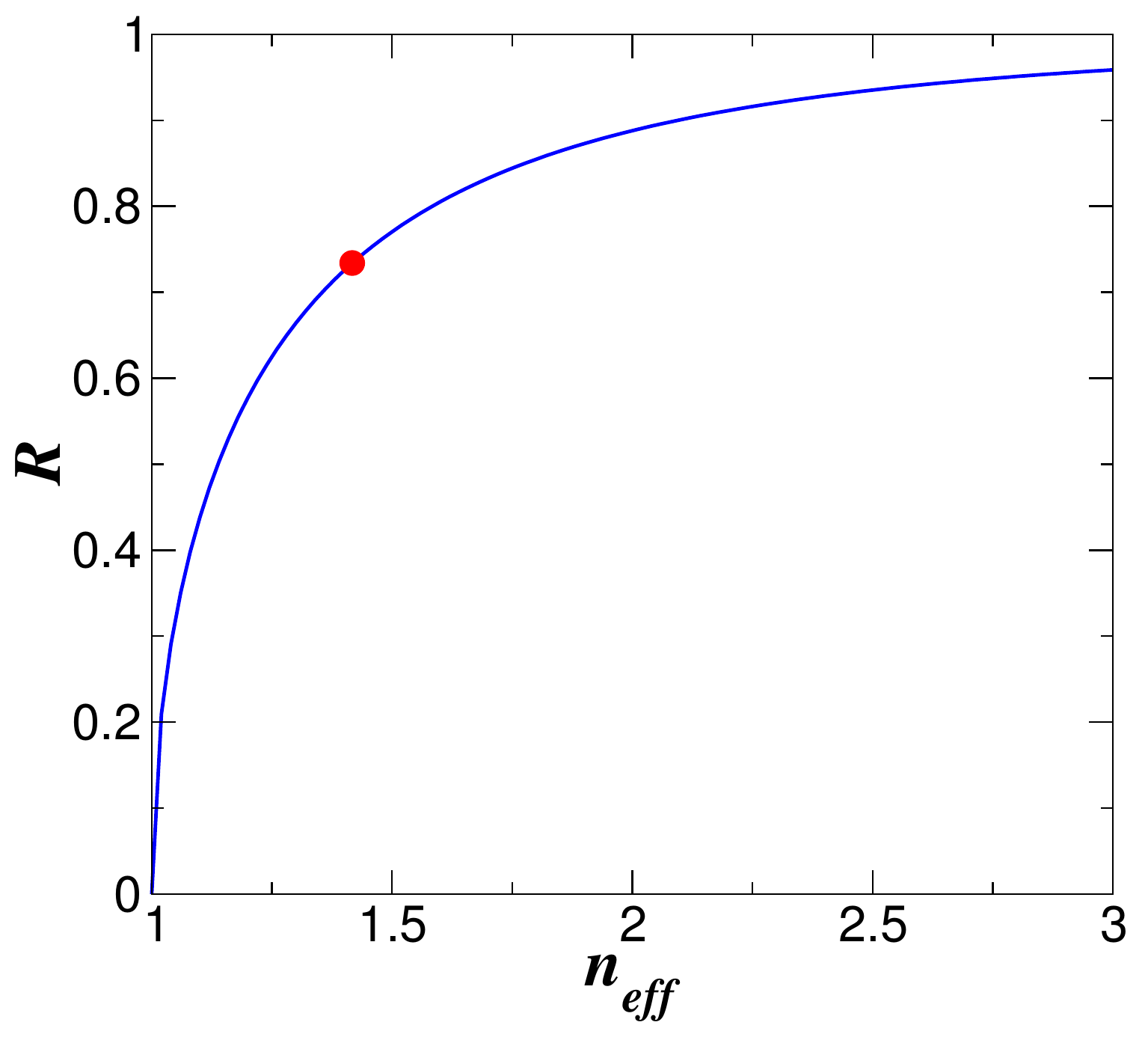}
\caption{\label{fig:av_refl} Angle and polarization averaged reflectivity
from a medium of refractive index $n_\text{eff}$ to vacuum.}\label{APavREfl}
\end{figure}
where R is the angle averaged reflectivity from the effective medium
of refractive index $n_\text{eff}$ to the vacuum. Considering a uniform angular distribution of the intensity, $R$ is given by
\begin{equation}
R=\mu_{c}+\int_{\mu_{c}}^{1}d\mu R\left(\mu\right)\textrm{,}\label{eq:R_20}
\end{equation}
where $\mu=\cos\left(\theta\right)$ ($\theta=$angle formed by the
propagation direction with the direction perpendicular to the slab),
$R\left(\mu\right)$ is the polarization averaged reflection at the
considered angle, and $\mu_{c}$ is the cosine of the critical angle
$\mu_{c}=\sqrt{1-1/n_\text{eff}^2}$. $R\left(\mu\right)$ is given by
\begin{equation}
R\left(\mu\right)=\frac{1}{2}\left|\frac{n_\text{eff}\mu-\mu_{2}}{n_\text{eff}\mu+\mu_{2}}\right|^{2}+\frac{1}{2}\left|\frac{n_\text{eff}\mu_{2}-\mu}{n_\text{eff}\mu_{2}+\mu}\right|^{2}\textrm{,}\label{eq:R_30}
\end{equation}
with $\mu_{2}=\sqrt{1+n_\text{eff}^2\left(\mu^{2}-1\right)}$ is the cosine
of the transmitted wave. In Figure~\ref{fig:av_refl}, the angle and polarization averaged
reflection $R$ is plotted as a function of the effective refractive
index. The averaged reflection at $n_\text{eff}=1.42$ ($R\simeq0.734$)
is highlighted with a full circle. 

\paragraph*{Evanescent wave tunneling around the gap center frequency $\nu^\prime_\text{Gap}$.} 
\begin{figure}[ht]
\centering
\includegraphics[width=.55\columnwidth]{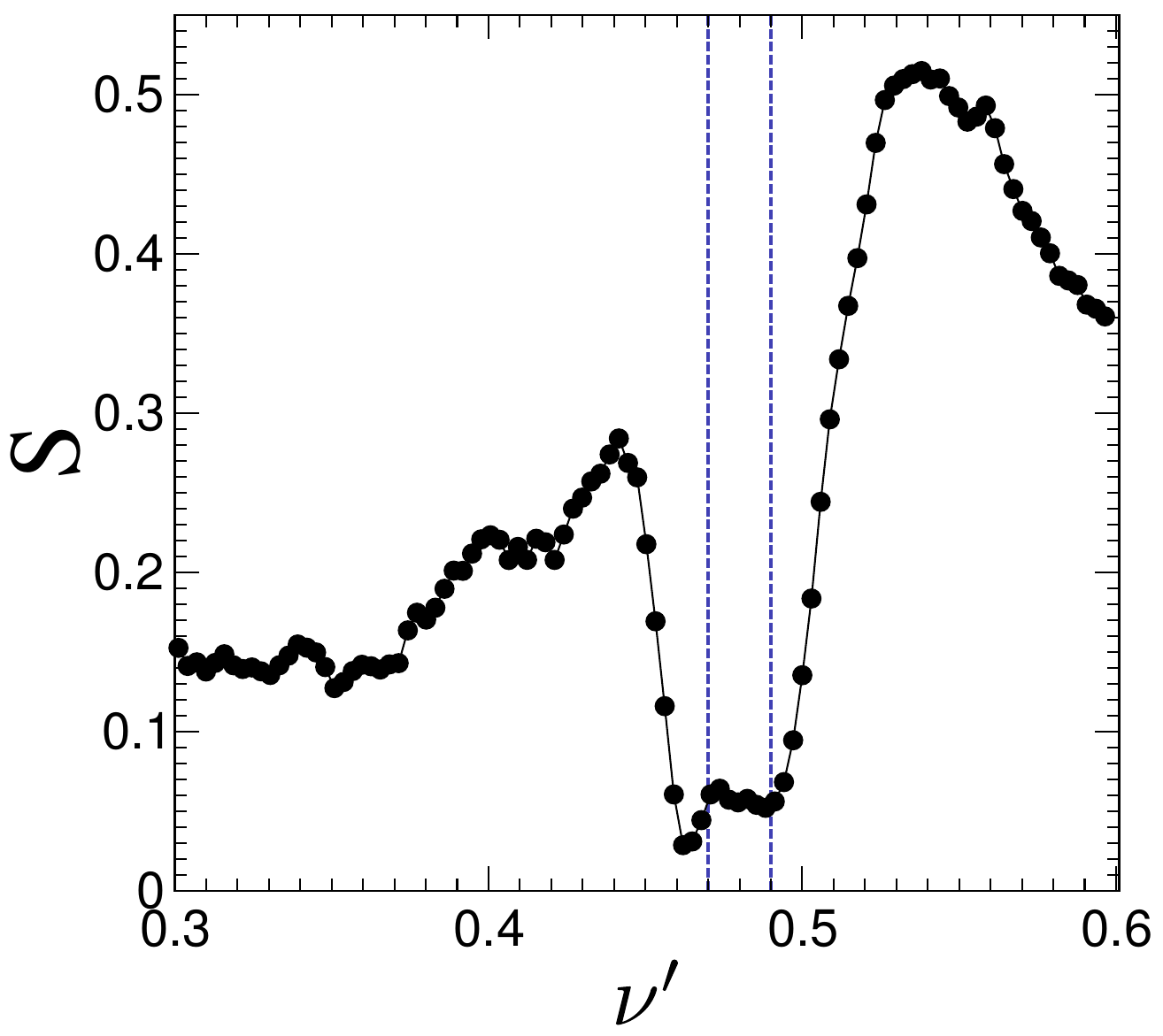}
\caption{\label{fig:PBG} Fitting error $S$ (least-squares) for a linear fit of $\ln T(L)$ over $L/a\in[0.3,2.4]$ (N=8 points). The initial exponential decay is observed for frequencies in the gap $\nu^\prime \in [0.46,0.49]$ and near the band edges. The vertical lines mark the band edges from MPB, as reported in Figure~\ref{fig:1} {(\bf{b})}.}
\end{figure}
Over a range of frequencies  around $\nu^\prime_\text{Gap}$ the transmittance $T(L/a,\nu^\prime)$ decays very rapidly, over the length scale of $L_B\sim a$, as shown in Fig.\ref{fig:1} {(\bf{d})}. We identify the center of this region as the photonic band gap (PBG), in agreement with the DOS calculations shown in Figure~\ref{fig:1} (b). To extract the decay length $L_B$ we perform a linear fit to the first eight points  $\ln(L/a)$ with $L/a\in[0.3,2.4]$ as shown in Fig.\ref{fig:1} {(\bf{d})}. In Figure
~\ref{fig:PBG} we show the corresponding error $S$ of the least-squares fit. We find that initial exponential decay for frequencies $\nu^\prime\in[0.46,0.495]$  
\newline \indent We note that for $T<10^{-2}$ we observe deviations from a single exponential 'evanescent' decay. With $L_B/a\simeq0.75$ at $\nu^\prime=0.48$ we would expect $T<10^{-3}$ for $L/a\ge5.5$. Our numerical data, shown for example in Fig.~\ref{fig:5}{(\bf{d})}, suggests $T>10^{-3}$ up to $L/a\simeq 10$. We tentatively explain this increased transmittance with the presence of spurious defect states at the boundary. When constructing the slab geometry with periodic boundary conditions, as discussed in the text, we expect a slight mismatch. This mismatch is due to the (to us) unknown exact size of the simulation box for the seed pattern, originally taken from reference \cite{song2008phase}. For the band gap calculations using MPB shown in Figure~\ref{fig:2}, we had to generate new and smaller disordered packings in order to match the available computational resources. To this end we were using the code provided by Torquato and coworkers upon request and described in \cite{skoge2006packing}. Here, the size of the simulation box is precisely known and the DOS in the gap is exactly zero. We note that in earlier work, we have observed that a slight mismatch of boundary conditions indeed leads to defect states in the band gap \cite{Froufe_PRL_2016}.  

\paragraph*{Diffuse transmission and reflection as a function of source position in SC theory.} 
We consider the Green function for the diffusion equation with a position-dependent
diffusion coefficient $D(z)$. In this section we normalize all lengths
and positions by the mean free path $\ell$, and the diffusion coefficient
$D\left(z\right)$ by the Boltzmann transport theory diffusion constant
\textbf{$D_{B}$}. 
\begin{equation}
-D\left(z\right)\frac{\partial^{2}g\left(z,z^\prime\right)}{\partial z^{2}}-\frac{\partial D\left(z\right)}{\partial z}\frac{\partial g\left(z,z^\prime\right)}{\partial z}=\delta\left(z-z^\prime\right)\label{eq:tr_10}
\end{equation}
with boundary conditions\begin{subequations}
\begin{align}
g\left(z=0,z^\prime\right) & =z_{0}D\left(z=0\right)\left.\frac{\partial g\left(z,z^\prime\right)}{\partial z}\right|_{z=0^{+}}\label{eq:tr_20a}\\
g\left(z=L,z^\prime\right) & =-z_{0}D\left(z=L\right)\left.\frac{\partial g\left(z,z^\prime\right)}{\partial z}\right|_{z=L^{-}}\label{eq:tr_20b}
\end{align}
\label{eq:tr_20}\end{subequations}

Assuming that $D\left(z\right)>0$, we can transform Eq.(\ref{eq:tr_10})
to
\begin{equation}
-\frac{\partial^{2}g\left(z,z^\prime\right)}{\partial z^{2}}-LN\left(z\right)\frac{\partial g\left(z,z^\prime\right)}{\partial z}=\frac{1}{D\left(z^\prime\right)}\delta\left(z-z^\prime\right)\textrm{,}\label{eq:tr_30}
\end{equation}
where
\begin{equation}
LN\left(z\right)\equiv\frac{d\left[\ln\left( D(z)\right)\right]}{dz}\label{eq:tr_40}
\end{equation}
For $z<z^\prime$ equation (\ref{eq:tr_30}) can be integrated. We have
\begin{align}
\frac{dg\left(z<z^\prime,z^\prime\right)}{dz} & =\frac{dg\left(z=0^{+},z^\prime\right)}{dz}\exp\left[-\int_{0}^{z}LN\left(x\right)dx\right]\nonumber \\
 & =\frac{dg\left(z=0^{+},z^\prime\right)}{dz}\exp\left[\ln\left(D\left(z=0\right)\right)-\ln\left(D\left(z\right)\right)\right]\nonumber \\
 & =\frac{dg\left(z=0^{+},z^\prime\right)}{dz}\frac{D\left(0\right)}{D\left(z\right)}\label{eq:tr_50}
\end{align}
using the boundary condition Eq.(\ref{eq:tr_20a}) we have
\begin{equation}
\frac{dg\left(z<z^\prime,z^\prime\right)}{dz}=\frac{g\left(0,z^\prime\right)}{z_{0}}\frac{1}{D\left(z\right)}\label{eq:tr_60}
\end{equation}
and, integrating Eq.(\ref{eq:tr_60}),
\begin{equation}
g\left(z<z^\prime,z^\prime\right)=g\left(0,z^\prime\right)\left[1+\frac{1}{z_{0}}\int_{0}^{z}\frac{1}{D\left(x\right)}dx\right]\textrm{.}\label{eq:tr_70}
\end{equation}
Analogously, for $z>z^\prime$, and considering the boundary condition given
by Eq.(\ref{eq:tr_20b}) we have
\begin{equation}
g\left(z>z^\prime,z^\prime\right)=g\left(L,z^\prime\right)\left[1+\frac{1}{z_{0}}\int_{z}^{L}\frac{1}{D\left(x\right)}dx\right]\label{eq:tr_80}
\end{equation}
To solve for the Green function Eq.(\ref{eq:tr_10}) we have consider
the discontinuity in the first derivative of $g\left(z,z^\prime\right)$
at $z\rightarrow z'^{\pm}$, i.e.
\begin{equation}
\frac{dg\left(z=z'^{+},z^\prime\right)}{dz}-\frac{dg\left(z=z'^{-},z^\prime\right)}{dz}=\frac{-1}{D\left(z^\prime\right)}\label{eq:tr_90}
\end{equation}
and the continuity of the Green function in one dimension
\begin{equation}
g\left(z=z'^{+},z^\prime\right)=g\left(z=z'^{-},z^\prime\right)\textrm{.}\label{eq:tr_100}
\end{equation}
Considering the solutions Eq.(\ref{eq:tr_70},\ref{eq:tr_80}) and
the previous matching conditions for the Green function Eq.(\ref{eq:tr_90},\ref{eq:tr_100})
we have\begin{subequations}
\begin{align}
g\left(0,z^\prime\right) & =z_{0}\frac{z_{0}+\int_{z^\prime}^{L}\frac{1}{D\left(x\right)}dx}{2z_{0}+\int_{0}^{L}\frac{1}{D\left(x\right)}dx}\label{eq:tr_110a}\\
g\left(L,z^\prime\right) & =z_{0}\frac{z_{0}+\int_{0}^{z^\prime }\frac{1}{D\left(x\right)}dx}{2z_{0}+\int_{0}^{L}\frac{1}{D\left(x\right)}dx}\label{eq:tr_110b}
\end{align}
\label{eq:tr_110}\end{subequations} If $J_{d}^{-}$ is the diffusive flux in $z<z^\prime$ and $J_{d}^{+}$is the diffusive flux in $z>z^\prime$ we can define the diffuse transmission and reflection coefficient $T_{SCT}\left(z^\prime\right)$ and $R_{SCT}\left(z^\prime\right)$ resp. as 
\begin{subequations}
\begin{align}
T_{SCT}\left(z^\prime\right) & \equiv J_{d}^{+}=-D\left(L\right)\frac{dg\left(z=L^{+},z^\prime\right)}{dz}=\frac{g\left(z=L^{+},z^\prime\right)}{z_{0}}=\frac{z_{0}+\int_{0}^{z^\prime}\frac{1}{D\left(x\right)}dx}{2z_{0}+\tilde{L}}\label{eq:tr_120a}\\
R_{SCT}\left(z^\prime\right) & \equiv-J_{d}^{-}=D\left(0\right)\frac{dg\left(z=0^{+},z^\prime\right)}{dz}=\frac{g\left(z=0^{+},z^\prime\right)}{z_{0}}=\frac{z_{0}+\int_{z^\prime}^{L}\frac{1}{D\left(x\right)}dx}{2z_{0}+\tilde{L}}\label{eq:tr_120b}
\end{align}
\label{eq:tr_120}\end{subequations}

Where we have defined
\begin{equation}\label{eq:lvzb}
\tilde{L}\equiv\int_{0}^{L}\frac{1}{D\left(x\right)}dx
\end{equation}

\clearpage

\end{document}